\documentclass[a4paper,11pt]{article}

\usepackage[a4paper,left=2.73cm,right=2.7cm,top=3cm,bottom=3.5cm]{geometry}
\usepackage{amsmath,amssymb,graphicx,caption,subcaption}
\usepackage[colorlinks=true,linktocpage=true,linkcolor=blue,citecolor=blue]{hyperref}
\usepackage[all]{xy}

\numberwithin{equation}{section}

\def\tl{{\tilde \Lambda}}
\def\tac{\tau}
\def\tension{T_b}
\def\cW{{\cal W}}
\def\cM{{\cal M}}

\begin{document}

\begin{titlepage}

\hfill{ICCUB-15-001}

\hfill{MPP-2015-2}

\vspace{1cm}
\begin{center}
{\huge{\bf Bifundamental Superfluids from Holography}}\\[5mm]

\vskip 45pt
{\large \bf Daniel Are\'an$^{\dagger}$ and Javier Tarr\'\i o$^{\sharp}$}

\vskip 20pt
{${\dagger}$ Max-Planck-Institut f\"ur Physik (Werner-Heisenberg-Institut),\\ F\"ohringer Ring 6, D-80805, Munich, Germany}\\[2mm]
{$\sharp$ Departament de F\'\i sica Fonamental and Institut de Ci\`encies del Cosmos,\\ Universitat de Barcelona, 
Mart\'\i\  i Franqu\`es 1, ES-08028, Barcelona, Spain.}\\

\vskip 20pt
{\texttt{darean@mpp.mpg.de\\[2mm] j.tarrio@ub.edu}}
\end{center}

 \vskip 10pt
\abstract{\normalsize
We  study the holographic dual of a $(2+1)$--dimensional 
s-wave superfluid that breaks an abelian $U(1)\times U(1)$ global symmetry group to the diagonal $U(1)_V$. The model is inspired by Sen's tachyonic action, and the operator that condenses transforms in the bifundamental representation of the symmetry group. 
We focus on two configurations: the first one describes a marginal operator, and the phase diagram at finite temperature 
contains a first or a second order phase transition, depending on the parameters that determine the theory. 
In the second model the operator is relevant and the finite temperature transitions are always second order.
In the latter case the conductivity for the current associated to the broken symmetry shows quasiparticle excitations at 
low temperatures, with mass given by the width of the superconducting gap. The suppression of spectral weight at low
frequencies is also observed in the conductivity associated to the conserved symmetry, for which the DC value 
decreases as the temperature is reduced.
}

\end{titlepage}

\tableofcontents

\hrulefill

\section{Introduction}

The realization of solutions with scalar hair in asymptotically AdS (aAdS) black holes in \cite{Gubser:2008zu}
prompted the study of strongly coupled superfluid systems via the holographic gauge/gravity duality.
The non trivial profile of the scalar field corresponds in the field theory to the condensation of an operator. 
If, in addition, the scalar is charged the model contains the breaking of the global symmetry dual to the gauge
symmetry in the bulk, hence superfluidity.

The seminal work \cite{Hartnoll:2008vx} considered an Einstein-Maxwell-Higgs (EMH) action, with the scalar field
coupled minimally to the $U(1)$ Maxwell field (see \cite{Hartnoll:2009sz,Horowitz:2010gk} for comprehensive reviews).
Further models with order parameters of different spin (s- \cite{Hartnoll:2008kx}, p- \cite{Gubser:2008wv,Jhsc}, 
and d-wave \cite{dwave}) have been developed
in the last years.
And while the initial setups in refs.~\cite{Hartnoll:2009sz,Horowitz:2010gk} are bottom-up models with the minimal field
content to describe the superfluid phase transition, microscopic embeddings have been proposed in the
framework of type IIB string theory \cite{Gubser:2009qm}, M-theory \cite{Gauntlett:2009dn}, and D-brane models \cite{Jhsc}.

In this work we draw intuition from string theory to modify the minimal setup of \cite{Hartnoll:2008vx}
and construct a model which realizes superfluidity as the spontaneous breaking of the ${U(1)\times U(1)}$ flavor
symmetry supported by a $D3\,$-$\overline{D3}$ brane pair in an asymptotically $AdS_4$  black hole geometry
($aAdS_4$ BH)\footnote{See
\cite{Jhsc} for models of holographic superfluids based on two overlapping probe branes.}.
Effective brane-antibrane actions derived from string theory \cite{Sen:2003tm} were first used in \cite{Casero:2007ae} to
realize chiral symmetry breaking through tachyon condensation in a stack of overlapping branes and antibranes (see also
\cite{tachQCD} for further developments in this direction). 
More precisely, those effective actions contain, in addition to the worldvolume gauge fields, a complex scalar, the tachyon,
which is the lightest open string mode extending between the brane and the antibrane.
In the models \cite{Casero:2007ae}-\cite{vQCD} the infrared region of a confining geometry
makes the tachyon condense and consequently break ${U(N_f)_L\times U(N_f)_R\to U(N_f)_V}$. 

We consider a setup corresponding to a single spacetime filling $D3\,$-$\overline{D3}$ pair in an $aAdS_4$ BH, 
and switch on a chemical
potential along the axial ${U(1)_A\subset U(1)\times U(1)}$ living on the worldvolume of the branes.
We couple the branes to gravity and take into account the backreaction, studying the system as the temperature is varied.
While at large temperature the setup is described by a solution with vanishing tachyon, below a certain temperature
the dominant solution contains a non trivial tachyon that realizes the spontaneous breaking ${U(1)\times U(1)\to U(1)_V}$,
needed for describing a superfluid phase transition.

This model possesses some characteristics that add novelty and interest to the analysis carried 
out in this article.
First, unlike most of the examples in the literature, in this setup the dynamics of both the scalar and the gauge field
are governed by a non-linear DBI term.\footnote{In \cite{Jing:2010zp,Gangopadhyay:2012am} holographic superfluids where
the dynamics of the gauge field is described by a Born-Infeld term were considered for the first time.}
Second, our setup is dual to a state with a non-zero density of matter
transforming in the fundamental representation of the gauge group, and the tachyon (corresponding to the order parameter
of the superfluid phase transition) is dual to a bifundamental operator. 
%Third, although still bottom-up,
%due to its brane motivation this system could serve as a toy model for more realistic \jt{realistic? meaning real world or stringy-origin?} holographic realizations of 
%superconductors. 
Third, our model allows  not only to tune the charge and the dimension of the operator that condenses
(dual to the charge and mass of the tachyon), but also to vary the strength of the backreaction of the matter sector
of the action. As a result a rich phase diagram is found. Finally, the presence of a second $U(1)$, the $U(1)_V$ not
coupled to the tachyon, opens up the possibility of studying unbalanced superconductors; as in 
\cite{Erdmenger:2011hp}-\cite{Amado:2013lia} one could switch on a chemical potential along the unbroken $U(1)_V$
and interpret the system as an unbalanced mixture.

The paper is organized as follows.
In section \ref{sec.model} we present the action of our model and derive the equations of motion.
Section \ref{sec.solutions} describes an analytic solution with vanishing tachyon, dual to the normal phase
of the system. There it is also shown that for some values of the parameters in the action, corresponding to different 
field theories on the boundary, the system
is unstable towards condensation of the tachyon. To end that section we discuss how to set up the numeric integration of
the condensed phase.
In section \ref{sec.masszero} we analyze the condensed phase for the case in which the scalar is dual to
a marginal operator.
We are able to identify the IR geometry of the zero temperature solution ($AdS_4$)
and to construct the phase diagram of the system. Both first and second order phase transitions are shown to 
occur, depending on the values of the parameters in the action.
The case of a scalar  dual to a relevant operator of dimension $\Delta=2$ is discussed in section \ref{sec.relevant}.
In this case the phase diagram presents only second order phase transitions.
Section \ref{sec.conducs} is dedicated to the study of the conductivity. Numerical results for the two theories
studied in previous sections are presented there.
We conclude in section \ref{sec.conclus} with a discussion of our results and possible continuations of this work.

\section{Setup}\label{sec.model}

In this section we present the model under consideration: Einstein gravity coupled to
the tachyonic DBI action describing a spacetime filling $D3$--$\overline{D3}$ brane pair.
The starting point is then Sen's tachyonic DBI action \cite{Sen:2003tm} (see also \cite{Sen:1999md}-\cite{Bergshoeff:2000dq})
\begin{equation}\label{eq.Sen}
S = - \int d^{4}x \, V \left(|\tac|\right) \left( \sqrt{ -  \det M^b} + \sqrt{- \det M^{\bar b}} \right) \ ,
\end{equation}
with $V$ a potential to be determined below and  matrices
\begin{equation}
M^{i}_{\mu\nu} = G_{\mu\nu} + 2 F_{\mu\nu}^{i} + D_{(\mu} \tac \overline{D_{\nu)}\tac} \ ,
\end{equation}
where $G_{\mu\nu}$ is the metric, $A^b$ ($A^{\bar b}$)  the $U(1)$ gauge field in the worldvolume of the (anti-)D-brane 
with field strength $F^i=dA^i$, and the complex scalar $\tac$ is the tachyon.\footnote{In the notation of
\cite{Garousi:2000tr} we have set $\pi \alpha'=1$ and  the transverse scalars to zero.
We follow conventions for (anti)symmetrized indices with no factors of $2$, \emph{i.e.}, $M_{(ab)}=M_{ab}+M_{ba}$, etc.} 
The tachyon is charged under the axial $U(1)_A$ subgroup of the original ${U(1)\times U(1)}$, and neutral under $U(1)_V$,
\begin{equation}
D_\mu = \partial_\mu + i\, q(A^b_\mu-A^{\bar b}_\mu) \ .
\end{equation}

We can now make explicit use of the combinations
\begin{equation}
\label{vecaxdef}
A^{V} = A^b + A^{\bar b}\,,\qquad A^{A} = A^b - A^{\bar b} \ ,
\end{equation}
defining the vectorial $U(1)_V$, and axial $U(1)_A$ gauge fields respectively. In terms of these one can write
\begin{subequations}
\begin{align}
M^{b}_{\mu\nu} & = G_{\mu\nu} +  F_{\mu\nu}^{V} +  F_{\mu\nu}^{A} + D_{(\mu} \tac \overline{D_{\nu)}\tac} \ , \\
M^{\bar b}_{\mu\nu} & = G_{\mu\nu} +  F_{\mu\nu}^{V} -  F_{\mu\nu}^{A} + D_{(\mu} \tac \overline{D_{\nu)}\tac} \ , \\
\label{covdev} D_\mu & = \partial_\mu + i q A^A_\mu  \ .
\end{align}
\end{subequations}

In the equations of motions obtained from \eqref{eq.Sen} it is consistent to set $A^V=0$, hence the 
action \eqref{eq.Sen} can be truncated to
\begin{equation}\label{eq.truncatedSen}
S = - 2 \int d^{4}x \,  V \left(|\tac|\right)  \sqrt{ -  \det \left[ G_{\mu\nu}  +  
F_{\mu\nu} + D_{(\mu} \tac \overline{D_{\nu)}\tac}  \right] }  \ ,
\end{equation}
where the only
gauge field is now the axial combination of the original ${U(1) \times U(1)}$  symmetry.

For our setup to be dual to a 3-dimensional system at finite temperature and chemical potential,
the $D3$--$\overline{D3}$ brane pair described by (\ref{eq.truncatedSen}) will be embedded in an  asymptotically $AdS_4$ 
black hole geometry.  
At low values of the temperature (measured with respect to the axial chemical potential) the expectation is that
the backreaction of the charged branes on the geometry becomes important \cite{Hartnoll:2009ns}.
To account for this fact we couple \eqref{eq.truncatedSen} to Einstein gravity, which
fixes the action of our model to be
\begin{equation}\label{eq.DBImodel}
S = \frac{1}{2\kappa^2} \int d^{4}x \left[ \sqrt{-G} \left( R - 2\tilde \Lambda \right)  - \tension  \, V(|\tac|) \sqrt{-\det \left[ {\cal P}[G]_{\mu\nu} +   F_{\mu\nu} +  \, D_{(\mu} \tac \overline{D_{\nu)} \tac}  \right]} \right] \ ,
\end{equation}
In this action $2\tl = - 6/\tilde L^2$ is a  cosmological constant and ${\cal P}[G]$
stands for the pullback of the metric 
onto the branes worldvolume (from now on we  consider the static gauge ${{\cal P}[G]_{\mu\nu}=G_{\mu\nu}}$).
Notice that we have written explicitly the tension of the brane $T_b$ in the Lagrangian. In our bottom-up model
we will consider it a tunable parameter which regulates the amount of backreaction by the branes.
As for the tachyon potential we will take it to be
\begin{equation}\label{eq.tacpotential}
V(\tac) = \exp \left( m^2 \, |\tac|^2 \right) \ ,
\end{equation}
which for small values of $\tau$ behaves as in \cite{Garousi:2000tr}, namely
\begin{equation}\label{eq.potentialexpansion}
V\sim 1 + m^2 \tac^*\tac + {\cal O}(\tac^*\tac)^2 \ ,
\end{equation}
where $m$ is then the mass of the tachyon, which is related to the dimension of the operator dual to $\tac$ in the 
standard way in holography, $\Delta(\Delta-3)=m^2 L^2$, with $L$ the radius of the asymptotic $AdS_4$ spacetime.

Notice that in this bottom-up approach we will take $m^2 L^2$ to be an order one parameter. 
As a consequence this model cannot be constructed from string theory without including
$\alpha'$ corrections \cite{tachQCD}.\footnote{Moreover, according to the original conjecture for the 
tachyonic action, the potential evaluated at its minimum should vanish \cite{Sen:1999md}.
Although this occurs for the confining geometries considered in \cite{Casero:2007ae}-\cite{vQCD}, it will not be true
for our superfluid solutions. More differences with a  top-down approach  consist of  the absence of a dilaton or
Wess-Zumino terms.}

Finally, notice that the radii appearing in $m^2 L^2$ and in the cosmological constant, $2\tl = - 6/\tilde L^2$,
are not the same.
To relate them consider $F_{\mu\nu}=\tac=0$ in \eqref{eq.DBImodel}, then the cosmological constant is shifted
\begin{equation}\label{eq.Lshift}
-2\tilde \Lambda \to - 2\tilde \Lambda - \tension \equiv -2 \Lambda = \frac{6}{L^2} \ .
\end{equation}
In the solutions of \eqref{eq.DBImodel} it is the shifted cosmological constant $\Lambda$ the one that 
controls the $AdS_{4}$ radius near the boundary.
In the following we will work in terms of the $AdS_4$ radius $L$ defined as in \eqref{eq.Lshift}.

\subsection{Ansatz and equations of motion}

We will be interested in static solutions with rotational and translational invariance, consequently
\begin{equation}
G_{\mu\nu} dx^\mu dx^\nu = G_{tt}(r) dt^2 + G_{rr}(r) dr^2 + G_{xx}(r) d\vec x_{2}^2 \ ,
\end{equation}
with negative $G_{tt}$.
Moreover, we will set the phase of the tachyon to zero, and consider solutions where the only non-vanishing component
of the gauge field is the temporal one:
\begin{equation}
\tac^*=\tac=\tac(r)\,,\qquad A_t=A_t(r)\,.
\end{equation}
Let us now write the equations of motion for this ansatz. First, it is useful to define
\begin{equation}\label{eq.curlw}
\cW = (G + F +  D_{(\mu}\tac \overline{D_{\nu)}\tac})^{-1} \equiv \cM^{-1} \ ,
\end{equation}
with components written with two risen indices: $W^{\alpha\beta}$. 
In particular we need
\begin{subequations}
\label{eq.wcomp}
\begin{align}
\Pi & =  \left[  A_t'^2 + \left( G_{tt} + 2 q^2  A_t^2 \tac^2 \right)  \left( G_{rr} + 2  \tac'^2 \right) 
\right] \ ,  \\
\cW^{tt} & =\frac{ G_{rr}+2 \tac'^2 }{\Pi } \ , \qquad
 \cW^{rr}  =  \frac{ 2q^2  A_t^2 \tac^2 + G_{tt} }{\Pi } \ , \\
 \cW^{[rt]} & =   \frac{- 2 A_t' }{\Pi }\ , \qquad\quad\,\,\,\,\,
\cW^{xx}  = G_{xx}^{-1} \ ,\\
\sqrt{\det \cM} & = \frac{G_{xx}\, \Pi}{\sqrt{G_{tt} + 2 q^2  A_t^2 \tac^2}\,\sqrt{G_{rr} + 2  \tac'^2}} \ .
\label{wrt}
\end{align}
\end{subequations}
Then, the equations of motion read
\begin{subequations}\label{eqs.fullequations}
\begin{align}
& R_{\mu\nu} - \frac{1}{2} G_{\mu\nu} R + \left( \Lambda - \tfrac{\tension}{2} \right) \, G_{\mu\nu} + \frac{\tension}{4} V(\tac) \frac{\sqrt{-\det \cM}}{\sqrt{-\det g}} \, G_{\alpha(\mu} G_{\nu)\beta} \cW^{\alpha\beta}  = 0 \ ,\label{einseqs}\\
& \frac{1}{V(\tac) \sqrt{-\det \cM}} \partial_r\left[ V(\tac) \sqrt{-\det \cM} \, \cW^{[rt]}  \right] + 4 \,  \cW^{tt}  \, q^2 \tac^2 A_t  = 0 \ , \label{gaugeq}\\
& \frac{1}{V(\tac) \sqrt{-\det \cM}} \partial_r \left[ V(\tac) \sqrt{-\det \cM} \, \cW^{rr} \, \tac' \right] -
\left( \frac{\partial_\tac V(\tac)}{2\,V(\tac)} +  \cW^{tt} q^2 A_t^2 \tac \right)  = 0 \ .\label{tacheq}
\end{align}
\end{subequations}

Next,we  partially fix diffeomorphism invariance by choosing as metric ansatz
\begin{equation}\label{eq.metricansatz}
ds^2  = -  g(r) e^{-\chi(r)} dt^2 + \frac{r^2}{L^2} d \vec x^2_{2} +  \frac{dr^2}{g(r)} \ ,
\end{equation}
and the equations of motion \eqref{eqs.fullequations} then reduce to the following set of two second order and two 
first order differential equations
\begin{subequations}\label{eqs.fullequationsans}
 \begin{align}
g'+ {g\over r}+\left( - 3 - \frac{\tension}{2} \right) \,r+{\tension\,r\,V(\tac)\over 2S}(1+2\,g\,\tac'^2)=0\ ,\\
  \chi'+{\,\tension\over S}\,r\,V(\tac)\,\tac'^2
  +{\,q^2\,\tension\over g^2\,S}\,r\,e^{\chi}\,V(\tac)\,A_t^2\,\tac^2=0\ ,\\
\tac''  -{\,g\,\over r Q}\,(r\,g\,\chi'-4Q-r\,g')\,\tac'^3+
  \left( -{2\,e^\chi\,g\,A_t'^2\over r\,Q}  +{2\over r}+{g'\over2}  \left({1\over g}+{1\over Q}\right)
  -{g\,\chi'\over 2Q}  \right)\tac' \\
+  {\,e^\chi\,A_t'^2\,V'(\tac)\over 2\,Q\,V(\tac)}
  +{q^2\,e^\chi\,A_t^2\,\tac\over g\,Q}-{\partial_\tac V(\tac)\over 2\,g\,V(\tac)}+\left(
  {2q^2\,e^\chi\,A_t^2\,\tac\over Q}-{\partial_\tac V(\tac)\over V(\tac)}\right)\tac'^2=0\ ,\nonumber\\
A_t'' + \left(
  {2(g-Q)\,\tac'\over Q\,\tac}+{2\over r}+{g'\over2}  \left({1\over g}-{1\over Q}\right)
  +{g\,\chi'\over 2Q}+{\,g\,\tac'^2\over r\,Q}\left(4Q+r\,g'-r\,g\,\chi'\right)
  \right)A_t'\\
  -{2q^2\,\,A_t\,\tac^2\over g}+{2(g-Q)\,A_t'^2\over A_t\,Q}+
  {g(Q-g) \over q^2\,\,r\,Q\,A_t^2\,\tac^2}\,A_t'^3-{4q^2}
  \,A_t\,\tac^2\,\tac'^2=0\ ,\nonumber
 \end{align}
\end{subequations}
where we have defined
\begin{subequations}
\begin{align}
 Q & = g-2q^2\,\,e^\chi\,A_t^2\,\tac^2\  , \\
 S & = \sqrt{1-e^\chi\,A_t'^2+2\,g\,\tac'^2-(2q^2\,/g)\,e^\chi\,A_t^2\,\tac^2(1 +2\,g\,\tac'^2)} \ .
\end{align}
\end{subequations}

As we will now see, in some limits these equations reduce to those of well know systems as the holographic superconductors
of \cite{Hartnoll:2008kx}.

\subsection*{Linear limit}

It is possible to recover the original holographic superconductor setup of \cite{Hartnoll:2008kx} in a limit in 
which the DBI term in the action is linearized.
This linearization is achieved by taking $T_b\to \infty$,  with
\begin{equation}
\frac{q}{\sqrt{T_b}} \to \tilde q \,\text{ fixed}\ , \qquad \sqrt{T_b}\, A \to \tilde A \,\text{ fixed}\ , \qquad \sqrt{T_b} \, \tac \to \tilde \tac \,\text{ fixed}\ .
\end{equation}
In this limit the action \eqref{eq.DBImodel} reduces to
\begin{equation}
S = \frac{1}{2\kappa^2} \int d^{4}x \sqrt{-G} \left(  R - 2 \Lambda - \frac{1}{4} \tilde F_{\mu\nu} \tilde F^{\mu\nu} 
- \tilde D_{\mu}\tilde \tac \tilde D^{\mu} \tilde \tac - m^2 \tilde \tac^2  \right) + {\cal{O}}(T_b^{-1}) \ ,
\end{equation}
where we have used the relation \eqref{eq.Lshift}. Indeed this action is that of the minimal model for holographic
superconductors constructed in \cite{Hartnoll:2008kx}.

\subsection*{Probe brane limit}

In the opposite limit, $\tension\to0$, the equations of motion for the metric decouple from the equations of motion for the 
$U(1)$ gauge field and the tachyon. Therefore, in this limit our system can be interpreted as a spacetime filling
$D3$--$\overline{D3}$ pair
in the Schwarzschild-$AdS_{4}$ spacetime, with the action describing the dynamics of the branes given by the DBI part of
\eqref{eq.DBImodel} alone.

\subsection*{Unbalanced superconductors}
In \cite{Bigazzi:2011ak}, a model realizing unbalanced holographic superconductors was constructed. In unbalanced
superconductors two different fermionic species with unbalanced populations contribute to the formation
of superconducting states. This imbalance, or chemical potential mismatch, can be described holographically by means
of a second $U(1)$ under which the order parameter is neutral \cite{Bigazzi:2011ak} (see also \cite{Erdmenger:2011hp}).

In our model, in addition to the chemical potential along the $U(1)_A$ directly coupled to the tachyon
(see eqs.~(\ref{vecaxdef}),(\ref{covdev})),
one could switch on a chemical potential along the remaining $U(1)_V$. The tachyon is not charged under this second $U(1)$,
and thus following \cite{Bigazzi:2011ak} one could treat it as dual to a chemical potential imbalance.
In \cite{Bigazzi:2011ak} the two $U(1)$s did not interact directly with each other, but only through their backreaction
on the geometry. This will be different in the present system due to the non-linearity of the DBI action.
Hence, it would be interesting to study the phenomenology of this model when switching on a chemical potential along
$U(1)_V$. We do not pursue this direction in this paper, but leave it instead for future works.

\section{Solutions}\label{sec.solutions}

In this section we present an analytic solution of the equations of motion which is dual to the normal phase of the system
(for which the charged scalar vanishes).
Next, we show that at low temperature this solution is unstable towards condensation of the scalar.
We study the dependence of this instability on the parameters of the model, which will be crucial
for the characterization of the phase diagram in the following sections.
Finally, we study the boundary conditions which will allow us to numerically construct the solutions with a non trivial
scalar (\emph{i.e.} those dual to the condensed phase).

\subsection{Normal phase solution}

A charged black hole solution to \eqref{eqs.fullequations} with $\tac=0$, corresponding to the normal phase of the dual 
theory, was found in \cite{Pal:2012zn}
\begin{subequations}\label{eq.normalphase}
\begin{align}\label{eq.blackening}
g(r) & = \frac{r^2}{L^2} \left( 1 +\frac{L^2\, \tension}{6} - \tension \, L^{4} |\rho| \frac{   \,{}_2F_1\left( - \frac{1}{2} , \frac{1}{4}  ; \frac{5}{4} ; - \frac{1}{\rho^2} \frac{r^{4}}{L^{4}} \right) }{ 2 \, r^{2}} - \frac{k_h}{r^3} \right) \ , \\
\chi & = 0 \ , \\
A_t' & = \rho \frac{L^{2}}{\sqrt{\rho^2 L^{4} + r^{4} } } \ ,
\end{align}\end{subequations}
where $\rho$ is a constant of integration related to the charge density of the black hole and $k_h$ is a constant of 
integration that can be translated into the radius of the horizon, $r_+$, by the condition $g(r_+)=0$ 
(with $r_+$ the largest root).

The thermodynamics of this solution was analyzed in \cite{Pal:2012zn,Tarrio:2013tta}, with temperature, entropy density, 
chemical potential, charge density and pressure density given respectively by
\begin{subequations}\label{eq.normalthermo}
\begin{align}
T & = \frac{r_+}{4\pi L^2} \left[ 3 + \tension\,L^2   \frac{ 1 - \frac{L^{2}}{r_+^{2}} \sqrt{\rho^2 + \frac{r_+^{4}}{L^{4}}} }{8\pi} \right] \ , \\
s & = \frac{2\pi}{\kappa^2}  \frac{r_+^{2}}{L^{2}} \ , \label{eq.entropy}\\
\mu & = \rho \frac{L^{2}}{r_+} \, {}_2F_1 \left( \frac{1}{2}, \frac{1}{4} ; \frac{5}{4} ; - \rho^2 \frac{L^{4}}{r_+^{4}}  \right)  \ , \\
Q & = \frac{T_b}{2\kappa^2} \, \rho \ , \\
P & = \frac{1}{2\kappa^2} \frac{1}{L^{4}} \left[ -k_h - \frac{L^{5}\, \tension}{4\sqrt{\pi}}|\rho|^{3/2} \Gamma\left( - {3/4} \right)  \Gamma\left( {5/4} \right) \right]  \ .
\end{align}
\end{subequations}

\subsection{Instabilities towards scalar condensation}\label{sec.instabilities}

We now want to study possible instabilities of the normal phase solution (\ref{eq.normalphase}) towards the condensation
of the tachyon. Let us then consider the fluctuation of the scalar field $\tau$ around that solution. 
This perturbation is described by the DBI part of the action \eqref{eq.DBImodel} with the metric and $U(1)$ field as 
in \eqref{eq.normalphase}.
It was shown in \cite{Seiberg:1999vs} that to describe fields appearing in the DBI action it is convenient to work
in terms of the open string metric, $s_{\mu\nu}$, defined as the inverse of the symmetric part of the 
$\cW(\tac=0)$ matrix \eqref{eq.curlw} (see \cite{Sen:1999md} for similar comments in the context of Sen's effective 
tachyonic action). 
Equivalently,
\begin{equation}
s_{\mu\nu} = G_{\mu\nu} - \left( F\cdot G^{-1}\cdot F \right)_{\mu\nu} \ .
\end{equation}
A second quantity that is important for the description of the perturbation is the effective running  coupling
\begin{equation}
\beta(r) = \frac{\sqrt{-\det s}}{\sqrt{-\det (G+F)}} \ ,
\end{equation}
that compensates the change of volume density from the original description to the one in terms of the open string metric.
For the solution \eqref{eq.normalphase}
\begin{equation}
\beta = \frac{r^{2}}{\sqrt{\rho^2 L^{4} + r^{4} } } \ , \qquad s_{\mu\nu}dx^\mu dx^\nu = \beta^2 \left( - g\, dt^2 
+ \frac{1}{g} dr^2 \right) + \frac{r^2}{L^2} d\vec x^{2}_{2} \ ,
\end{equation}
with $g(r)$ given in \eqref{eq.blackening}.
The equation of motion for the perturbation of the scalar field then reads
\begin{equation}\label{eq.KGOSM}
\frac{\beta}{\sqrt{-s}} \partial_\mu \left[ \frac{1}{\beta} \sqrt{-s} \, s^{\mu\nu} \partial_\nu \tac \right] 
- \left(m^2 + q^2 A_t^2 s^{tt} \right) \tac=0 \ .
\end{equation}
Close to the boundary $A_t\to \mu$, and \eqref{eq.KGOSM} describes a free scalar in $AdS_{4}$ with mass squared 
given by $m^2$.

\subsection*{Instability at extremality from the $AdS_2$ throat}

\begin{figure}[t]
\begin{center}
\begin{subfigure}[b]{0.47\textwidth}
\includegraphics[width=\textwidth]{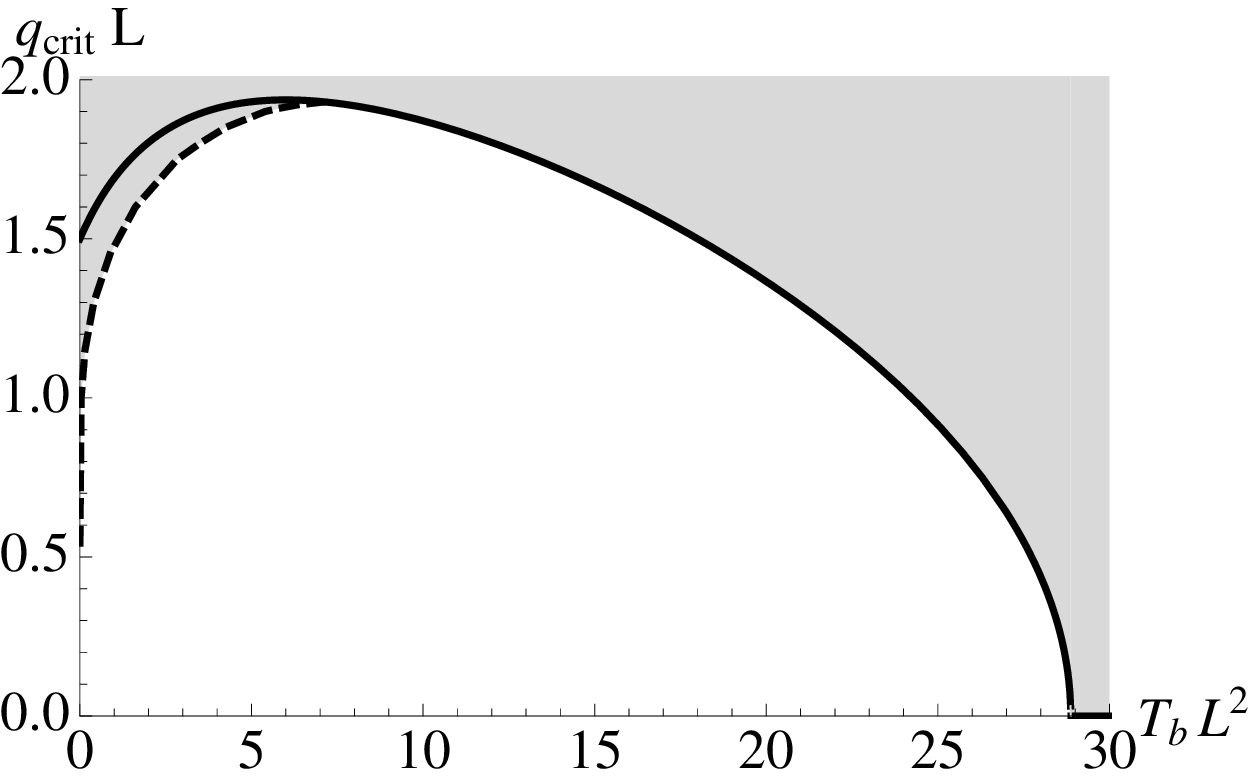}
\caption{$m^2 L^2=-2$.\label{fig.BFboundmmin2}}
\end{subfigure}
~
\begin{subfigure}[b]{0.47\textwidth}
\includegraphics[width=\textwidth]{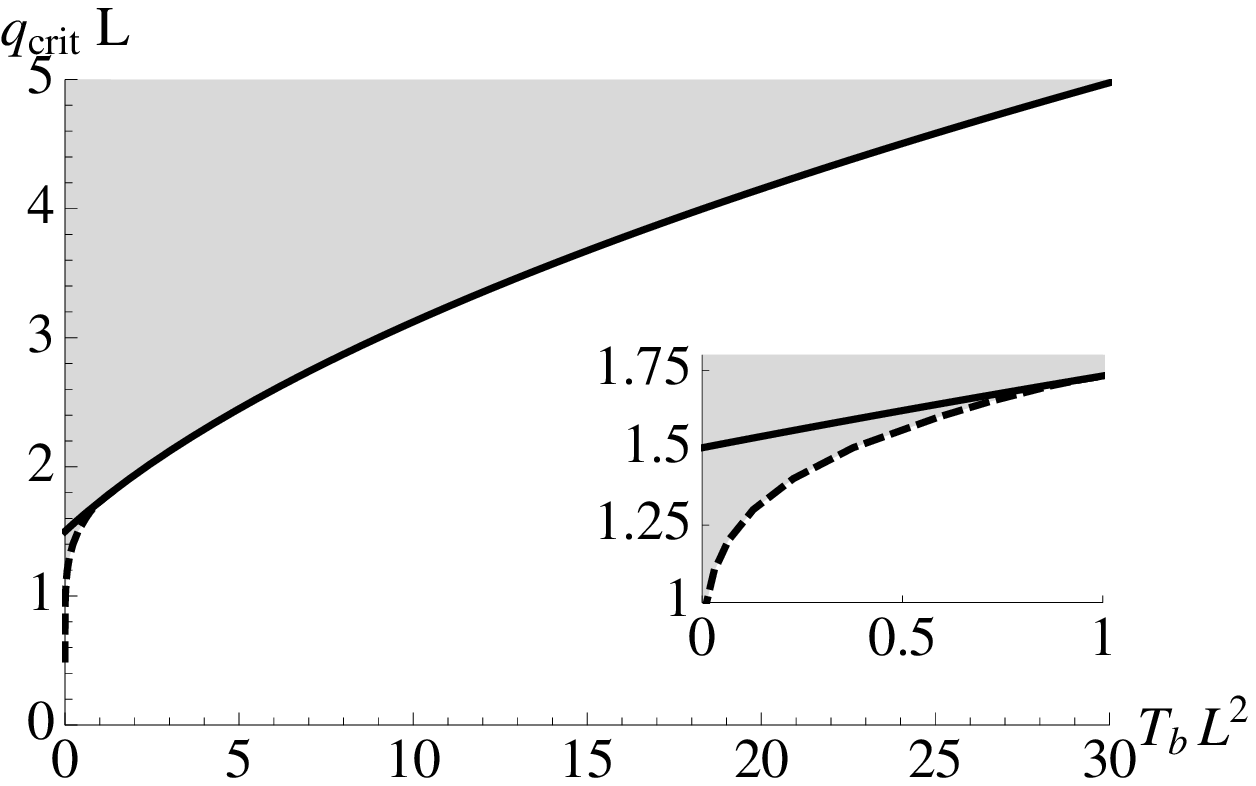}
\caption{$m^2 L^2=0$.\label{fig.BFboundm0}}
\end{subfigure}
\caption{Critical value of the charge as given by $m_{\rm eff}^2 L_{AdS_2}^2 =-1/4$ (solid line) and by the onset of the  
instability in the a$AdS_4$ region (dashed line) for two values of the scalar mass. The shadowed area is unstable towards 
condensation of the scalar field.} \label{fig.BFbound}
\end{center}
\end{figure}

At zero temperature the horizon is a double root of the blackening function \eqref{eq.blackening}.
Therefore, close to the extremal horizon (at $r=r_0$) both $\beta$ and $A_t^2 s^{tt}$ approach a constant.
Hence, in the zero temperature limit the equation of motion (\ref{eq.KGOSM}) describes a scalar with effective mass
\begin{equation}
m_{\rm eff}^2 = \lim_{r\to r_0} \left(m^2 + q^2 A_t^2 s^{tt}\right) = m^2 -2 q^2 \frac{6 + \tension \, L^2}{\tension^2L^4}  \ ,
\end{equation}
in the near-horizon geometry $AdS_2\times \mathbb{R}^{2}$, with $AdS_2$ radius
\begin{equation}
L_{AdS_2}^2 = \frac{ L^2}{6  }\, \frac{2\,\tension^2 L^4}{\left( 6 + \tension \, L^2 \right) \left( 6 + 2 \, \tension\, L^2 \right)} \ .
\end{equation}
As was first shown in  \cite{Denef:2009tD,Horowitz:2009ij}, there is an instability when the effective mass violates
the Breitenlohner-Freedman (BF) bound of $AdS_2$: ${m_{\rm eff}^2 L_{AdS_2}^2 <-1/4}$.
The violation of the BF bound defines a critical charge $q_{\rm crit}$, dependent on the mass of the scalar field and 
the tension parameter $\tension$, above which the scalar $\tac$ is unstable.
In figure \ref{fig.BFbound} we plot this critical value as a function of the tension $T_b$ both for the theory 
with $m^2\,L^2=-2$ and $m^2\,L^2=0$. This results in a curve (solid line) above which the normal phase
described by eq. \eqref{eq.normalphase} is unstable towards condensation of the scalar.

From the arguments in \cite{Kaplan:2009kr,Iqbal:2010eh}  the instability caused by the violation of the 
BF bound of the $AdS_2$ near-horizon region gives rise to a continuous phase transition. 
In particular, near the extremal horizon $r_0$ the equation of motion \eqref{eq.KGOSM} for the scalar field takes
the form
\begin{equation}
\partial_r \left( (r-r_0)^2 \tac' \right) - m_{\rm eff}^2 L_{AdS_2}^2 \tac = 0 \ ,
\end{equation}
which, for $m_{\rm eff}^2L_{AdS_2}^2<-1/4$ has as solution
\begin{equation}\label{eq.BKTscalar}
\tac = \frac{c_\tac}{\sqrt{r-r_0}} \sin \left[ \sqrt{-\left( \frac{1}{4} + m_{\rm eff}^2 L_{AdS_2}^2 \right) } 
\log \frac{r-r_0}{r_{\text UV}-r_0}  \right] \ ,
\end{equation}
where $r_{\text UV}$ is a UV scale at which the $AdS_2$ region has to be corrected to recover the $AdS_4$ asymptotics of
the full solution, and $c_\tac$ is a normalization.

As explained in \cite{Kaplan:2009kr,Iqbal:2010eh}, equation \eqref{eq.BKTscalar} implies that there is a quantum phase 
transition if we allow the tension of the brane to vary from a BF bound-violating to a non-BF bound-violating value. 
For tensions slightly into the unstable region (\emph{i.e.}, $m_{\rm eff}^2 L_{AdS_2}^2\lesssim -1/4$) the critical 
temperature at which the phase transition takes place and the value of the condensate at vanishing temperature are given by
\begin{equation}\label{eq.BKT}
T_c \sim \Lambda_{\text UV} \, \exp \left[ \frac{-\pi}{\sqrt{-\left( \frac{1}{4} + m_{\rm eff}^2 L_{AdS_2}^2 \right) }} 
\right] \ , \quad \langle {\cal O}_\tac \rangle \sim \Lambda_{\text IR} \, \exp \left[ \frac{-\pi}{2\sqrt{-\left( \frac{1}{4} 
+ m_{\rm eff}^2 L_{AdS_2}^2 \right) }} \right]  \ ,
\end{equation}
with $\Lambda_{\text IR}$ ($\Lambda_{\text UV}$) an IR (UV) scale.

\subsection*{Instability at extremality from the boundary of $AdS_2$}

Let us now look again at figure \ref{fig.BFbound}, which illustrates the conclusions of the stability analysis
carried out in the previous subsection.
That analysis established that for points in the $(T_b,q)$ plane above the critical curve ${m_{\rm eff}^2\,L_{AdS_2}^2=-1/4}$
(solid line), the normal phase solution \eqref{eq.normalphase} is unstable at small temperatures. 
For points below that curve the BF bound is not violated. In the linear model of holographic superfluidity 
\cite{Denef:2009tD} this region is stable and dominates the large temperature part of the phase diagram.
However, in the present case a careful analysis shows that this no longer holds; for low values of the tension there is 
a region below the critical curve in which the $AdS_2$ BF bound is not violated, and yet there is an instability of
the scalar, indicating the preference for a  condensed phase. This is the region between the dashed and the solid lines in
figure \ref{fig.BFbound}.

To understand this new instability let us focus first on the probe approximation, $\tension=0$, at zero temperature.
The system becomes that of a probe brane in $AdS_{4}$, with an electric field on the worldvolume of the branes turned on.
The open string metric describing the fluctuation of the scalar governs the form of the equation of motion for $\tac$. 
Near the origin this equation becomes
\begin{equation}\label{eq.AdS4inst}
\tac''+\frac{2}{r} \tac' + \frac{q^2\,L^2}{r^2} \tac = 0 \ ,
\end{equation}
which is the  equation of a scalar in $AdS_2$ with effective mass $m_{\rm eff}^2\,L^2=-q^2\,L^2$  (in units of the $AdS_{2}$ 
radius).\footnote{One should keep in mind that, although effectively the scalar $\tac$ behaves near the horizon as if it were
in an $AdS_2$ geometry, the system corresponds to a probe in $AdS_4$.}
The  solution of this equation reads
\begin{equation}
\tac = \frac{c_\tac}{\sqrt{r}} \sin \left[ \sqrt{-\left( \frac{1}{4} - q^2\,L^2 \right) } \log \frac{r}{r_{\text UV}} \right] \ ,
\end{equation}
As one can see, there is a
minimum value of the charge for this solution to exist, be real, and well behaved at the origin.
Consequently, at $T_b\to 0$ there is an instability of the normal phase solution for
\begin{equation}
q\,L\geq \frac{1}{2} \ ,
\end{equation}
independently of the mass of the scalar (and the dimensionality of the theory).
Interestingly, as we explain below, we have found that this extremal instability for $q\,L>1/2$ extends to small values of 
$T_b$.
However, at non-zero $T_b$, the instability is not anymore caused by a violation of the BF bound in the  
$AdS_2$ throat. It must then be related to the boundary conditions associated to the UV of the throat, \emph{i.e.}, the 
region where the throat opens up to the asymptotically $AdS_4$ geometry. In particular this is an effect of the 
non-linearities of our model, since it is not present in setups like  \cite{Denef:2009tD}.

 These generic boundary conditions allow for the unstable modes always present in $AdS_2$ to survive in the spectrum.
 A similar scenario is given by the 
% Notice the similarity with the
 %This is similar to 
 instabilities studied in a different setup in \cite{Faulkner:2010gj}, where the authors allow mixed boundary conditions 
 for the scalar field, as opposed to the Dirichlet boundary conditions that we impose in the $aAdS_4$ boundary in the 
 present paper.

We have been able to find the values of the charge at which, for a given tension $T_b$, the $T=0$ instability
we have just described occurs.
The result is given by the dashed line in figure \ref{fig.BFbound}.
This line is constructed by finding the normalizable solution to \eqref{eq.KGOSM} at several small temperatures and 
extrapolating down to $T=0$.
As we will see later, at very small temperatures the phase transitions are always second order, which guarantees that 
this prescription to find  the dashed line in figure \ref{fig.BFbound} gives the correct result.\footnote{If the 
transitions were first order we would have found the value at which metastable solutions cease to exist.}

\subsection{IR boundary conditions of the condensed phase}
\label{sec.backreaction}
Once we have shown that the normal phase solution is unstable towards condensation of the tachyon the next step
is to find the solutions with non trivial scalar, which are dual to the condensed phase of the system.
To construct such a  solution we have to integrate the equations of motion numerically.
We can make use of the following scaling symmetries:
\begin{equation*}
\begin{tabular}{ccc|cccc}
$t$ & $x^i$ & $r$ & $g$ & $e^\chi$ & $A_t$ & $L$ \\
\hline\\[-4mm]
$e^{\alpha}$ & $e^{\beta}$ & $e^{\gamma}$ & $e^{2\gamma}$ & $e^{2(\alpha+\gamma)}$ & $e^{-\alpha}$ & $e^{\beta+\gamma}$
\end{tabular}
\end{equation*}

We use the $\gamma$-scaling  to impose that the metric  approaches  $AdS_4$ with unit radius when $r\to\infty$.\footnote{
We  do so in the remaining of this paper, although  we  restore the  $AdS_4$ radius for the dimensionless combinations 
$\tension\,L^2$, $q\,L$ and $m^2\,L^2$ for presentation purposes.}
There are  two remaining scaling symmetries which leave $L$ invariant
\begin{subequations}
\begin{align}\label{eq.chisymmetry}
& t \to  e^\alpha \, t  \ , \qquad  \chi \to \chi + 2\alpha  \ , \qquad A_t \to  e^{-\alpha } \, A_t  \ , \\
&x^i \to e^\beta \, x^i \ , \qquad r \to e^{-\beta} r \ , \qquad g \to e^{-2\beta} \, g \ , \qquad   \chi \to \chi 
- 2\beta \ .\label{eq.rsymmetry}
\end{align}
\end{subequations}
We use \eqref{eq.rsymmetry} to fix the position of the horizon at $r_+=1$.
The symmetry \eqref{eq.chisymmetry} will be used once the numeric integration has been performed to relate the periodicity 
of time in the UV with the temperature.

The asymptotic behavior of the fields in the IR is given by a series expansion whose leading terms at finite temperature are
\begin{subequations}\label{eq.IRasymps}
\begin{align}
g & = e^{\chi_+/2} 4\pi \overline{T} (r-r_+) + \cdots \ , \\
A_t & = a_+ (r-r_+) + \cdots \ , \\
\chi & = \chi_+ + \cdots \ , \\
\tac & = \tac_+ + e^{-\chi_+/2}\frac{1-a_+ e^{\chi_+} m^2}{4\pi \overline{T}}  \frac{\tac_+}{r_+} (r-r_+) + \cdots \ ,
\end{align}
\end{subequations}
where $\chi_+$, $a_+$ and $\tac_+$ are unspecified constants and
\begin{equation}\label{eq.horizontemD}
\overline{T} ={g'\,e^{-{\chi\over2}}\over4\pi}\bigg\vert_{r=r_+}
= e^{-\chi_+/2}\frac{ \sqrt{1-a_+^2 e^{\chi_+}}  
\left( 6+\tension  \right)-  e^{m^2 \tau_+^2} \tension }{2 \sqrt{1-a_+^2 e^{\chi_+}} } \ ,
\end{equation}
is related to the temperature of the configuration.
The reason why $\overline{T}$ is not readily identified with the temperature is that the appropriate boundary condition 
is given by the  periodicity of the euclidean time on the boundary.
Actually, near the UV we  impose that $g\to r^2$ and $\chi \to 0$ to calculate the temperature, but this cannot be done 
until the full numerical solution is obtained.

 The strategy we follow to calculate the  numerical solutions is as follows.
From eq.~\eqref{eq.IRasymps} we have three free parameters in the IR: $\chi_+$, $a_+$, and $\tau_+$.
While on the boundary we want to impose two boundary conditions:
%On the boundary there are have two boundary conditions: 
the vanishing of the non-normalizable mode of the scalar, and
$\chi\to0$. We will fix $\chi_+=1$ from the beginning, and make use of the scaling symmetry \eqref{eq.chisymmetry} once the 
numeric solution is obtained
to ensure that $\chi\to0$ at the boundary. Then, we expect  a family of solutions given by a relation between $\tau_+$ 
and $a_+$ that ensures the correct boundary condition for $\tau$ at infinity. The resulting one-parameter family of
solutions will give us the value of the condensate as a function of the temperature of the system.

\subsection{Free energy and VEVs}

Before presenting our solutions let us explain how we can detect a phase transition and calculate the condensate of the 
scalar operator dual to $\tac$.

When two or more solutions coexist at the same value of temperature and chemical potential we should determine which 
one  is thermodynamically favored by comparing their free energies.
This quantity is obtained holographically from the evaluation of the euclidean on-shell action on the boundary.
From equation \eqref{einseqs} it follows that the euclidean on-shell action is
\begin{equation}\label{eq.Ionshell}
I_\text{on-shell}=\frac{\beta_{\overline{T}}\, V_{2}}{\kappa^2} \int dr \sqrt{g} \, R^x{_{x}} =\frac{\beta_{\overline{T}}\, V_{2}}{\kappa^2} \int dr \, \partial_r\left[ - e^{-\chi/2}\,r \, g \right] =\frac{\beta_{\overline{T}}\, V_{2}}{\kappa^2} \left[ - e^{-\chi/2}\,r \, g \right]^{r_\Lambda}_{r_+} \ ,
\end{equation}
where we have integrated up to a cutoff $r_\Lambda$, $V_2$ is the spatial volume of $\mathbb{R}^2$ and 
$\beta_{\overline{T}}=1/\overline{T}$ the period of the euclidean time in the UV.
The lower limit of this expression vanishes since $g\to0$ there, whereas the upper limit diverges when 
$r_\Lambda\to\infty$, and must be regulated with a set of counterterms evaluated at the surface $r=r_\Lambda$
\begin{equation}\label{eq.Icounterterm}
I_\text{counterterms} = \frac{\beta_{\overline{T}}\, V_{2}}{\kappa^2} \sqrt{\gamma} \left( K - 2 - \frac{\tension}{2} \tac^2 \right)_{r=r_\Lambda} \ ,
\end{equation}
where $\gamma$ is the induced metric on the constant-radius slice, and $K=K_{ab}\gamma^{ab}$ with 
$K_{ab}=\nabla_{(a}n_{b)}$ the  extrinsic curvature and $n_b$ the components of an outward-pointing unit vector 
orthogonal to the constant-radius slice
\begin{equation}\label{eq.BYtensor}
K_{ab} = \mathrm{diag}\left( \frac{1}{2}e^{-\chi}\sqrt{g} \left( g'-g\,\chi' \right) , \, r\,\sqrt{g} ,\, r\, \sqrt{g}\right) \ , \qquad K = \frac{r\,g'+g\,(4-r\,\chi')}{2\,r\,\sqrt{g}} \ .
\end{equation}
The first term in \eqref{eq.Icounterterm} corresponds to the Gibbons-Hawking term, and is needed for the definition of the 
variational problem in gravity, the second term is a volume counterterm that cancels the divergent behavior of the 
asymptotic $AdS_4$ spacetime  \cite{Balasubramanian:1999re}, and the last term is needed to cancel a divergence given 
by the scalar field for some values of its mass.\footnote{And in particular when it is the leading term near the boundary 
the one that is kept fixed in the variational problem.}
A term proportional to $\sqrt{\gamma}\, \tau^3$ would produce a finite contribution to the free energy, but since the 
original action is even in the scalar we do not include it.

The internal energy and pressure can be obtained as conserved charges from the Brown-York tensor  
\cite{Balasubramanian:1999re}, and in our case reduce to the $tt$ and $xx$ components of
\begin{equation}
T_{ab} = -\frac{1}{\kappa^2} \frac{\sqrt{\gamma_{xx}^2} }{\sqrt{\gamma_{tt}}} \left( K_{ab} 
- K \gamma_{ab} + \left( 2 + \frac{\tension}{2} \tac^2 \right) \gamma_{ab} \right) \ .
\end{equation}

Finally, from the action \eqref{eq.DBImodel} complemented with the counterterm \eqref{eq.Icounterterm} we can calculate the 
variation of the on-shell action
with respect to the non-normalizable mode of the scalar field, $\tac_0$. This variation is dual to the vacuum expectation 
value (VEV) of the corresponding operator
\begin{equation}\label{eq.condensate}
\langle {\cal O}_\tac \rangle \equiv T \frac{\delta I_\text{on-shell}}{\delta \tau_{0}} = 
- T \frac{\tension}{2\kappa^2} \beta_{\overline{T}} \, V_2 \left( V(\tac)\, \sqrt{\det \cM} \, \cW^{rr}\tac' 
+ 2 \sqrt{\gamma} \, \tac  \right)  \frac{\delta \tac}{\delta \tau_{0}} \Bigg|_{r\to\infty} \ .
\end{equation}
In the following sections we will write down explicit versions of this expression and that
of the free energy
for the two cases analyzed in this paper (namely the $m^2L^2=-2$ and $m^2=0$ theories).

\section{Phase diagram for the model with a marginal operator}\label{sec.masszero}

In this section we  discuss the case in which the scalar field is dual to a marginal operator with conformal 
dimension $\Delta=3$. Holographically this is obtained by taking the mass of the scalar to zero, $m^2=0$. 
Then, the potential given in eq. \eqref{eq.tacpotential} is just $V=1$.

In order to read the physical observables of the dual theory we need to compute the boundary expansions of the
different fields in our setup.
For a massless scalar the two leading terms of the independent solutions to the scalar equation  near the boundary 
($r\to\infty$) are
\begin{equation}
\tac  \simeq  \tau_0 + \cdots  + \frac{\tau_3}{r^3} +  \cdots \ .
\end{equation}
The source of the dual scalar operator is determined by $\tau_0$ and the VEV is given in terms of $\tau_3$.
We are interested in the normalizable solution with zero source, $\tau_0=0$, and  we  impose this  in the following.

With $\tau_0=0$ the remaining equations of motion can be solved for large $r$ as
\begin{subequations}\label{eq.uv}
\begin{align}
g & \simeq r^2 \left[ 1 + \frac{{\cal E}}{r^3} + \cdots \right] \ , \\
\chi & \simeq \chi_0 +   \frac{3\, \tension}{2}  \frac{\tac_3^2}{r^6} +  \cdots \ , \label{eq.chiuvm0}\\
A_t & \simeq \bar \mu + \frac{\bar Q}{r} + \cdots \ .\label{eq.atuvm0}
\end{align}
\end{subequations}
The free energy ($\Omega=-T \left(I_\text{on-shell}+I_\text{counterterm}\right)$), internal energy ($E$), and pressure ($P$)
obtained from these UV ($r\to\infty$) expansions, and the relations \eqref{eq.Ionshell} -
\eqref{eq.BYtensor} are given by
\begin{equation}\label{eq.freeenergy}
\Omega  = V_2 \frac{e^{-\chi_0/2}   }{2\kappa^2} {\cal E} = -e^{-\chi_0}P = -e^{-\chi_0}\frac{E}{2}\ .
\end{equation}

As we discussed in the previous section, when we integrate the equations of motion numerically
we fix $\chi=1$ at the horizon, and use the rescaling \eqref{eq.chisymmetry} to meet the boundary condition
$\chi(r\to\infty)=0$. 
Applying this rescaling, the temperature
and the chemical potential of the dual theory are given by
\begin{equation}
T = e^{\chi_0/2} \overline{T} \ , \qquad \mu = e^{\chi_0/2} \bar \mu  \ ,
\end{equation}
in terms of $\overline{T}$, $\bar\mu$, and $\chi_0$ defined in eqs.~\eqref{eq.horizontemD}, \eqref{eq.atuvm0} and 
\eqref{eq.chiuvm0} respectively.
We then use the chemical potential to construct dimensionless ratios of the physical quantities in our problem, and
present our results in terms of these.
For example, 
the corresponding ratios for the temperature and free energy read
\begin{equation}\label{eq.numericexpressions}
\frac{T}{\mu} = \frac{\overline{T}}{\bar \mu} \ , \qquad \frac{2\kappa_2}{V_2} \frac{\Omega}{\mu^3} = e^{-2\chi_0} \frac{{\cal E}}{\bar \mu^3} \ ,
\end{equation}
where the left-hand side of these expressions corresponds to field theory quantities and the right-hand side to parameters 
from the numeric solution.

From equation \eqref{eq.condensate} we obtain that the condensate density of the marginal operator can be read from the 
UV asymptotic solution as
\begin{equation}\label{eq.condensatemarginal}
2\kappa^2 \frac{\langle {\cal O}_\tac \rangle}{V_2} = 3\, \tension \,  \tac_3 \ .
\end{equation}

\begin{figure}[tb]
\begin{center}
\includegraphics[width=0.8\textwidth]{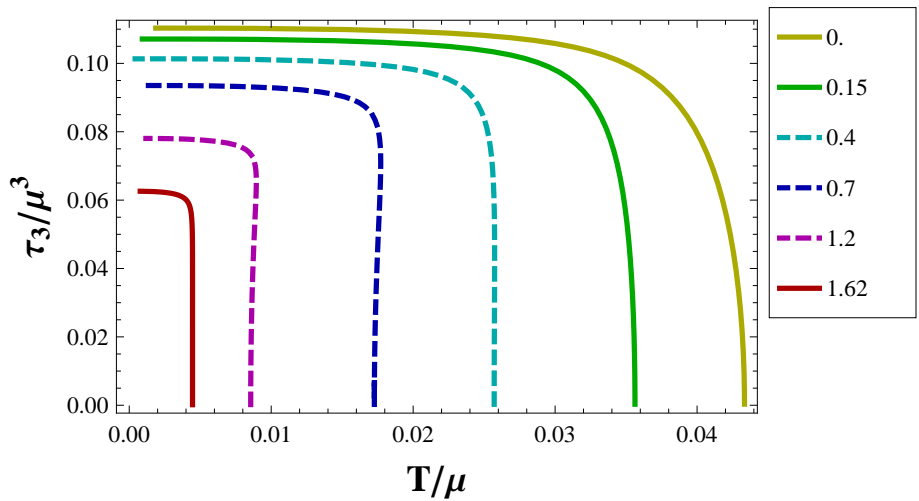}
\caption{(Color online) Values of the condensate  as a function of the temperature for $m^2\,L^2=0$ and $q\,L=5/2$. 
Each curve corresponds to a different value of $\tension\,L^2$ (see inset). 
Dashed (solid) lines indicate first (second) order phase transitions.
\label{fig.changeinordercondensate}}
\end{center}
\end{figure}

In figure \ref{fig.changeinordercondensate} we plot the value of the condensate (actually $\tac_3$, which is related 
to the condensate by means of eq.~\eqref{eq.condensatemarginal}) as a function of the temperature for a single value of 
the scalar charge $q\,L=5/2$, and different values of the tension, $T_b$.
The outer lines correspond to phase transitions with  larger critical temperature and vice versa.
For the solid lines a fixed value of the temperature corresponds unambiguously to a value of the scalar condensate. 
At the critical temperature the value of the condensate is turned on continuously, and the phase transition is of second 
order.
For the dashed lines there is a region around the critical temperature in which metastable configurations exist for a 
given temperature. 
For $T>T_c$ the thermodynamically preferred phase is the one with no condensate, and for $T<T_c$ the one with the 
largest condensate. 
There is a finite jump in the value of  $\langle {\cal O}_3 \rangle$ as the temperature is lowered, and the phase 
transition is first order.

\begin{figure}[tb]
\begin{center}
\begin{subfigure}[b]{0.475\textwidth}
\includegraphics[width=\textwidth]{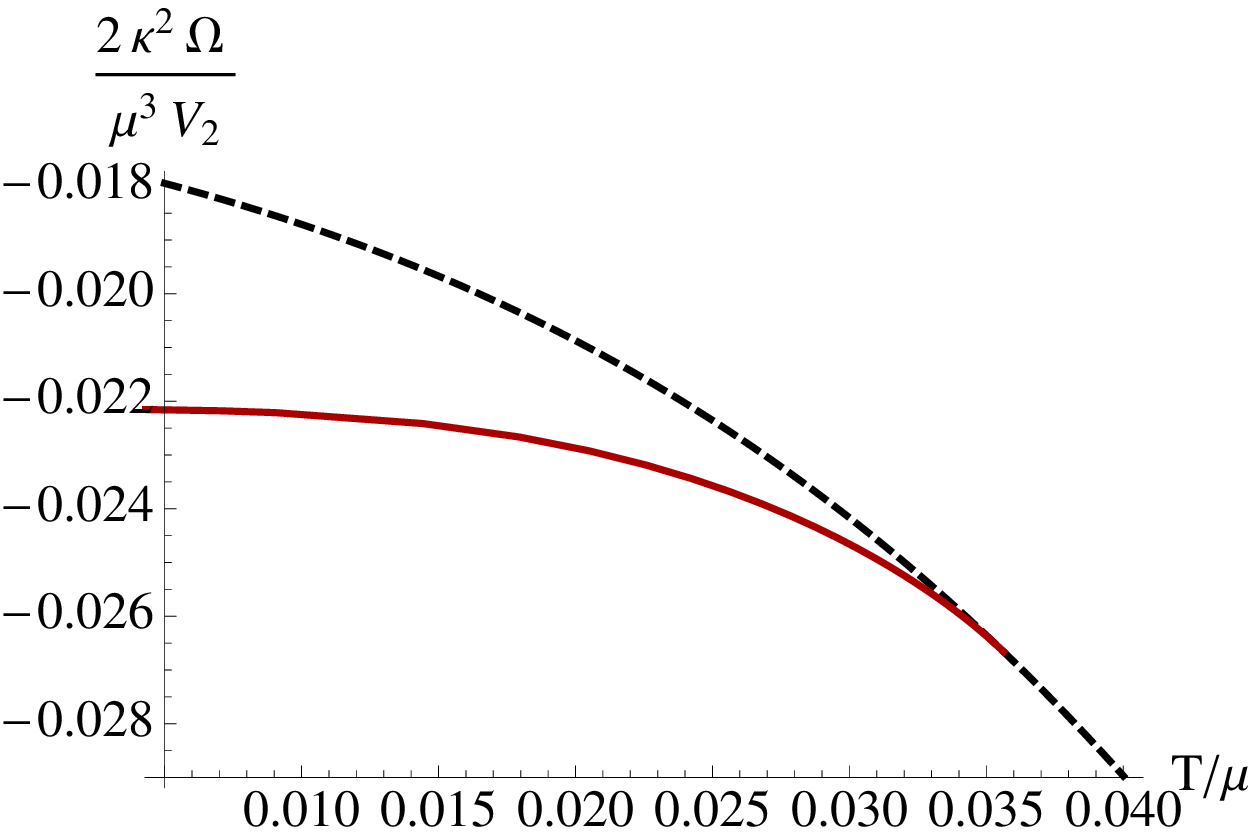}
\caption{$\tension \,L^2 = 0.15$ (second order) \label{fig.secondorderchangeinorder}}
\end{subfigure}
~~~
\begin{subfigure}[b]{0.475\textwidth}
\includegraphics[width=\textwidth]{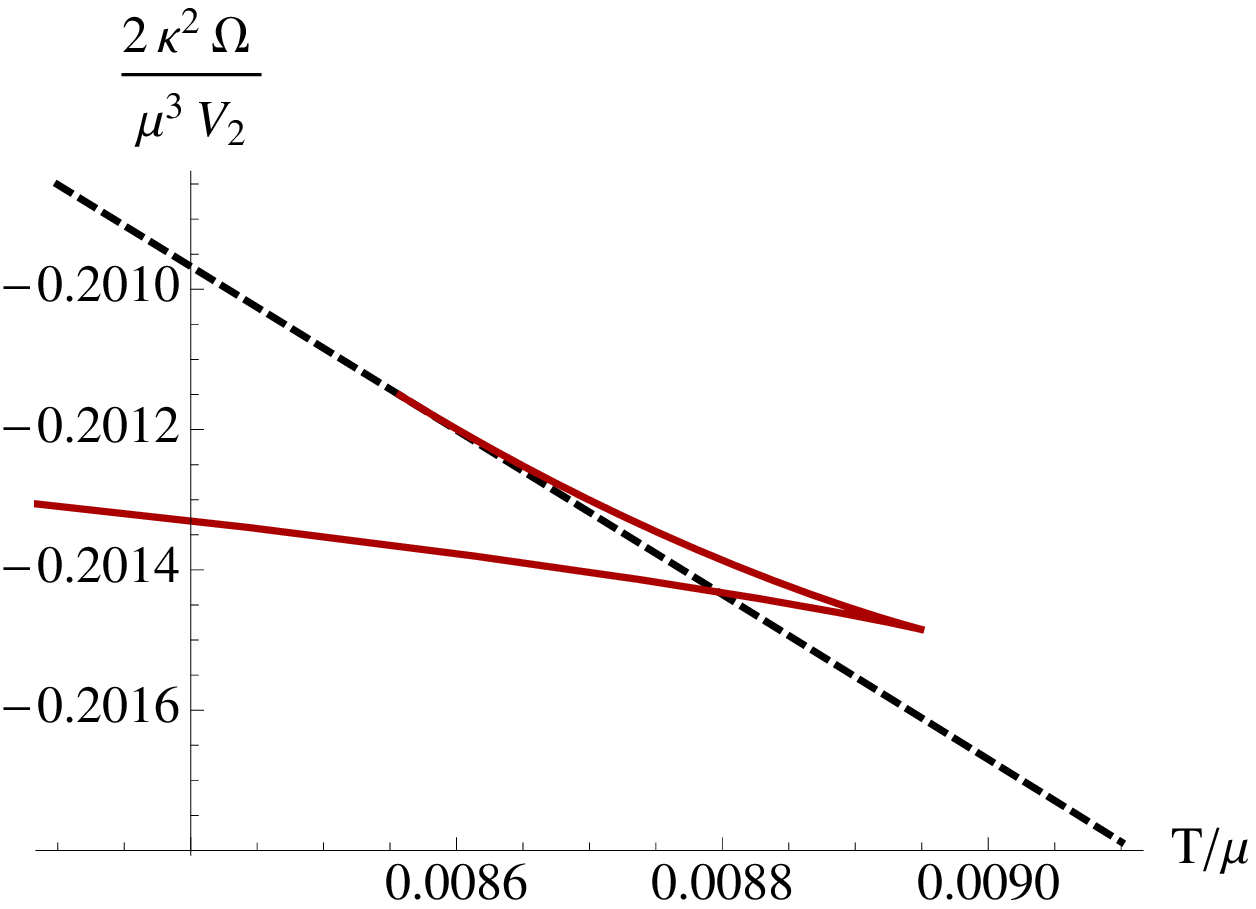}
\caption{$\tension \,L^2 = 1.2$ (first order)\label{fig.firstorderchangeinorder}}
\end{subfigure}
\caption{Plots of the  the free energy in the normal (black, dashed line) and condensed phase (red, solid line) as a function of the temperature for $m^2\,L^2=0$ and $q\,L=5/2$.} \label{fig.changesinorder}
\end{center}
\end{figure}
The value of the critical temperature is determined by studying the free energy of the system,
and in figure \ref{fig.changesinorder} we present two examples of this calculation.  
In both graphs the line corresponding to the free energy of the normal phase was obtained from eq.\eqref{eq.normalthermo} 
as $\Omega=-P$.
The left-hand side graph, figure \ref{fig.secondorderchangeinorder}, corresponds to a second order transition,
while the plot on the right-hand side, figure \ref{fig.firstorderchangeinorder}, corresponds to a first order one: 
the derivative of the free energy with respect to the temperature presents a finite jump.

\begin{figure}[tb]
\begin{center}
\includegraphics[width=0.5\textwidth]{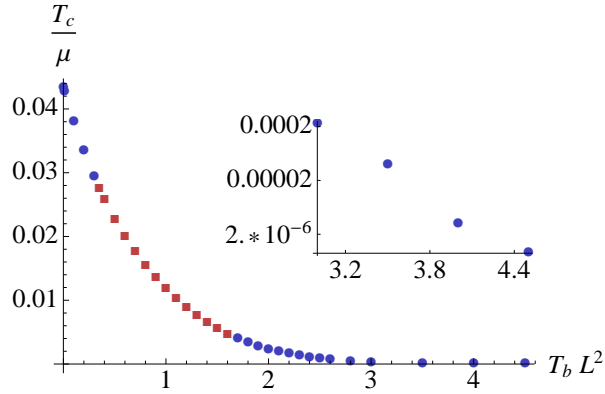}
\caption{Phase diagram for $m^2 L^2=0$ and $q\, L=5/2$. Blue disks correspond to second order phase transitions and red 
boxes to first order ones. The inset is a zoom on the rightmost points. Beyond $\tension={16/3}$ there is no condensed 
phase at any temperature.} \label{fig.PDmass0q2p5}
\end{center}
\end{figure}
\begin{figure}[tb]
\begin{center}
\includegraphics[width=0.5\textwidth]{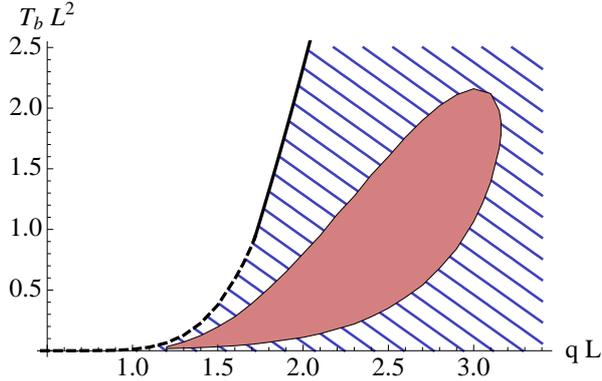}
\caption{Order of the phase transition in the $q\,L$ -- $\tension\,L^2$ plane. The red, shadowed region denotes values 
of the parameters where a first order phase transition occurs, and the blue striped region surrounding it to values where
a second order phase transition does. The thick solid and dashed lines are  the curves  in figure \ref{fig.BFboundm0}. 
In the white region there is no condensed phase.} \label{fig.limittensions}
\end{center}
\end{figure}

In figure \ref{fig.PDmass0q2p5} we plot the value of the critical temperature as a function of the tension $T_b$
for the same value  of the charge as before, namely $q\, L=5/2$. To ease the identification of the phase transition order
for a given tension, 
we have indicated the second (first) order phase transitions with blue circles (red squares).
As is clear from the plot, for this value of the charge there exists a region, corresponding to intermediate values of the 
tension, for which the phase transition is of the first order kind. 
However, this region does not extend to arbitrary values of the charge $q\, L$, as one can see in figure 
\ref{fig.limittensions}. In this plot we present a scan of the parameter space $(T_b,q)$, showing that only for a
closed region in that plane do first order phase transitions occur.
At zero values of the tension, or for values of the parameters where the critical temperature is infinitesimally small,
 the phase transition is always second order.
Pictorially, the red, shadowed region in figure \ref{fig.limittensions} is fully surrounded by values of the parameters 
where the transition is second order.

An interesting observation about the plot in figure \ref{fig.limittensions} is that 
the solid thick line 
separating the region with no condensate (large values of the tension) and the region with second order phase transitions 
(striped region) corresponds to an holographic BKT quantum phase transition governed by the instability occurring  at 
the $AdS_2$ throat described in section \ref{sec.instabilities}. Indeed, for the marginal operator under consideration 
this instability takes place along the curve
\begin{equation}\label{eq.maxtension}
\tension^c \,L^2  = \frac{4\,q^2L^2 -9}{3} \ ,
\end{equation}
where we have defined $\tension^c\,L^2$ as the tension that saturates the $AdS_2$ BF bound at extremality as a function of 
the charge of the scalar field.
On the other hand, the dashed line in figure \ref{fig.limittensions} corresponds to the instability beyond the $AdS_2$ throat,
also discussed in section \ref{sec.instabilities}, and describes a second order quantum phase transition.
These two lines (solid and dashed) are the same ones as those in figure \ref{fig.BFbound}.
To understand better these phase transitions it is worth studying in detail the behavior of the system at zero temperature. 
Although this study will be the subject of the next section, let us point out that in the inset of
figure \ref{fig.PDmass0q2p5} it is shown how the critical temperature approaches zero exponentially fast when 
$\tension\,L^2 \to \tension^c \, L^2 =16/3$, following the behavior in equation \eqref{eq.BKT}.

\begin{figure}[tb]
\begin{center}
\begin{subfigure}[b]{0.45\textwidth}
\includegraphics[width=\textwidth]{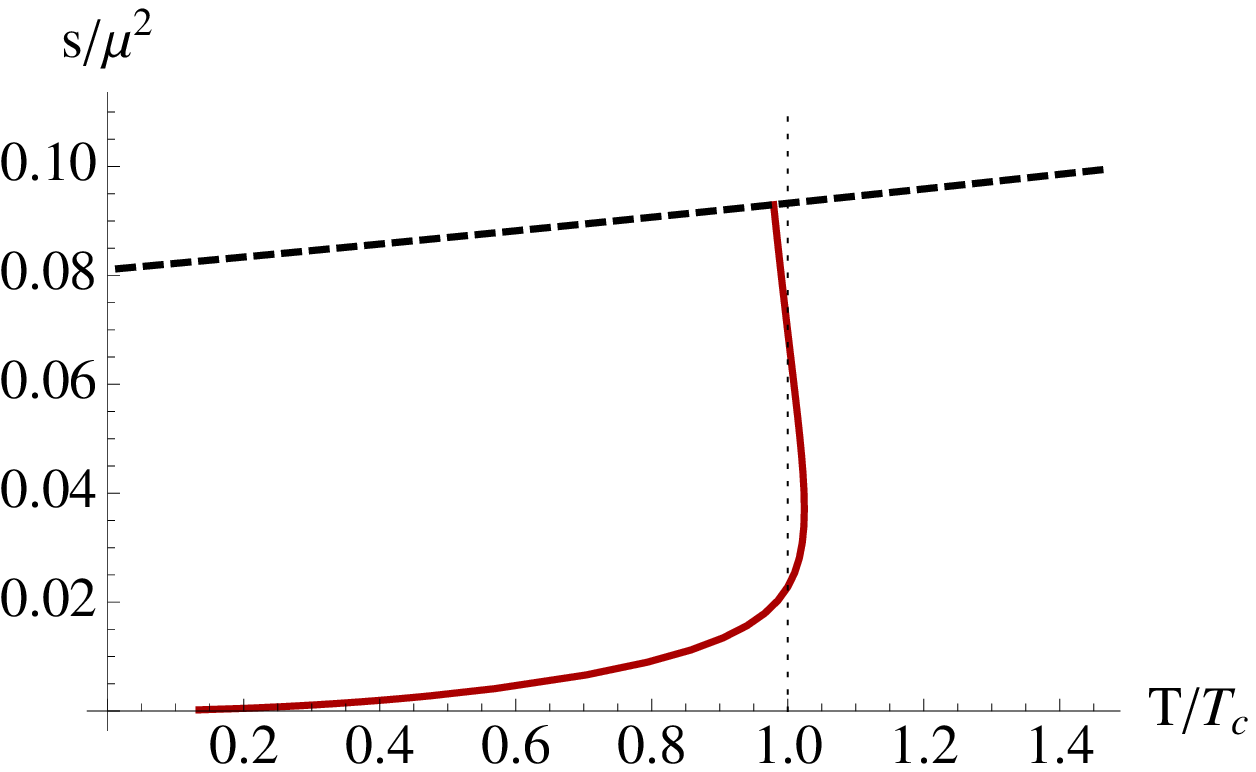}
\caption{$\tension \, L^2=1.2$.}
\end{subfigure}
~
\begin{subfigure}[b]{0.45\textwidth}
\includegraphics[width=\textwidth]{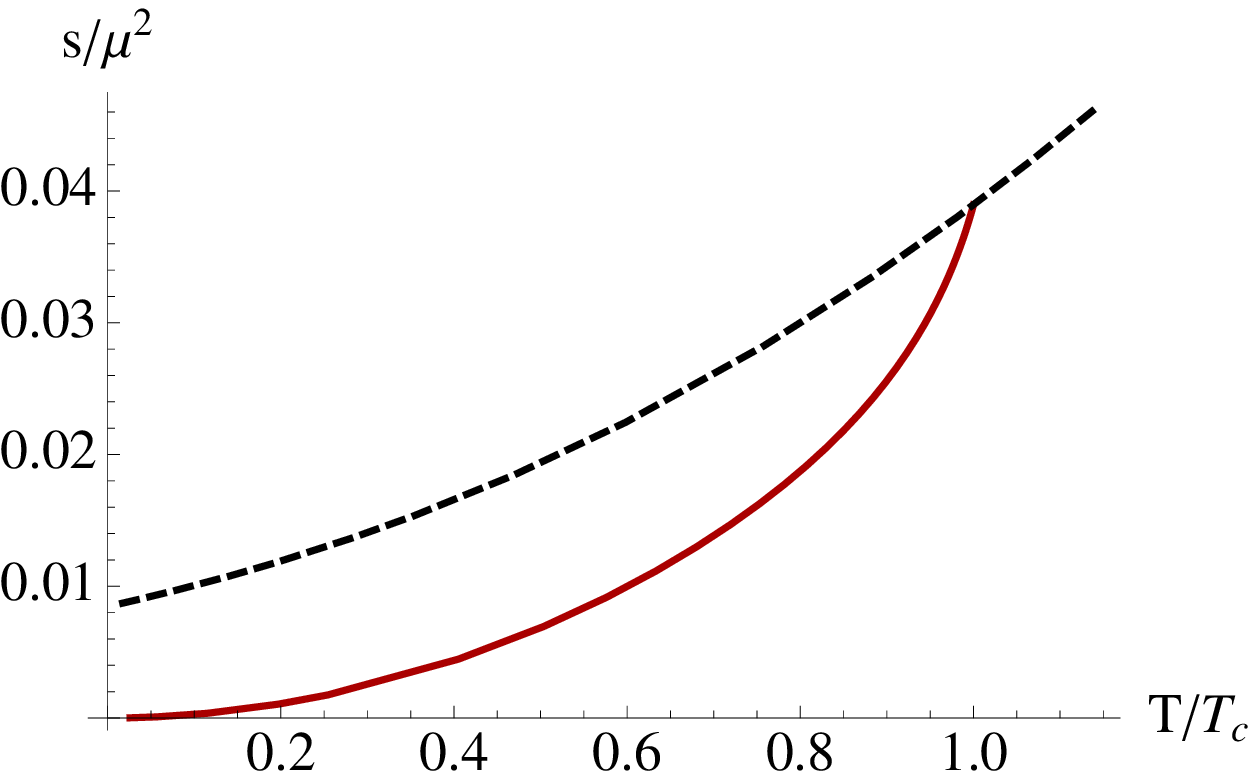}
\caption{$\tension \, L^2=0.15$.}
\end{subfigure}
\caption{Plots of entropy density (in units of $\kappa^2/(2\pi)$) as a function of the temperature at $m^2\,L^2=0$, 
$q\,L=2.5$ and two different values of the tension corresponding to (a) $1^{st}$ order and (b) $2^{nd}$ order phase 
transitions.} \label{fig.m0entropies}
\end{center}
\end{figure}

We  finish this section with 
some representative plots of the behavior of the entropy density as a function of the temperature in figure 
\ref{fig.m0entropies}. 
Both in the normal and broken phases the entropy density is given by the horizon radius squared ($r_+^2$) up to some 
proportionality constants (see eq. \eqref{eq.entropy}). In our numerical integration we keep $r_+=1$ by making use of the 
scaling symmetry \eqref{eq.rsymmetry}, hence the rescaling invariant ratio $s/\mu^2$ is given by
\begin{equation}
{s\over\mu^2}\,{\kappa^2\over2\pi}={e^{-\chi_0}\over\bar\mu^2}\ ,
\end{equation}
where $\bar\mu$ and $\chi_0$ are respectively the asymptotic values of $A_t$ and $\chi$ defined in eq. 
\eqref{eq.uv}.
In the plots in figure \ref{fig.m0entropies} the dashed, black line corresponds to the normal phase, and it goes to a 
constant in the zero temperature limit. The continuous, red line corresponds to the condensed phase; near $T=0$ it vanishes 
quadratically with the temperature.

\subsection{Quantum phase transitions}
\label{subsec.qpt}

In this section we will study the zero temperature limit of the system with a massless scalar, which will allow us to 
characterize better the quantum phase transitions described above.

The zero temperature geometry corresponding to the ground state of the condensed phase consists of a domain wall
interpolating between two $AdS_4$ with the  same radius. We construct it by integrating numerically from the IR
($r\to0$) towards the UV($r\to\infty$).
In the IR
the metric is  $aAdS_4$ with the same $AdS$ radius as in the UV, constant scalar field, and vanishing gauge field. 
Conformal symmetry is therefore recovered in the IR end point of the RG flow for the condensed 
phase.\footnote{The same phenomenon occurs in the Einstein-Maxwell-scalar model of \cite{Horowitz:2009ij}, whose treatment
we follow closely in this section (see also \cite{Gubser:2009cg}).}
The asymptotic solution around the origin of spacetime is, for $q\neq0$ (recall that we are setting $L=1$),
\begin{subequations}\label{eq.zeroTexpansion}
\begin{align}
g & \simeq r^2 \left( 1-\frac{\tension}{4}e^{\chi_0}(\alpha+1) \, r^{2 \alpha} + {\cal O}\left( r^{4 \alpha} \right) \right) \ , \\
\chi & \simeq \chi_0-\frac{\tension}{4}e^{\chi_0} \frac{(\alpha+1)(\alpha+2)}{\alpha} \, r^{2 \alpha} 
+ {\cal O}\left( r^{4 \alpha} \right)  \ , \\
A_t & \simeq r^{1-\alpha} \left( r^{2 \alpha} + {\cal O}\left( r^{4 \alpha} \right) \right) \ , \label{eqatzeroTIR}\\
\tac & \simeq \frac{\sqrt{(\alpha+1)(\alpha+2)}}{\sqrt{2}\, q} \left( 1-\frac{q^2}{4} \frac{e^{\chi_0}}{\alpha\, (3+2\alpha)} \, r^{2 \alpha} + {\cal O}\left( r^{4\alpha } \right) \right) \ , 
\end{align}
\end{subequations} 
where $\alpha>0$ is a free parameter.

In \eqref{eq.zeroTexpansion} we have imposed four conditions by choosing the IR behavior of the fields, and we need to set
two further boundary conditions in the UV ($r\to\infty$). These are as before $\chi=0$, and the vanishing of the source term
in the UV asymptotics of the scalar.
Again, we can make use of the scaling symmetry \eqref{eq.chisymmetry} to make $\chi\to0$ as $r\to\infty$.
Then, by tuning the constant $\alpha$ in our IR solutions we ensure the vanishing of the source term in the UV of the scalar.

In the probe limit $\tension\,L^2=0$, the equations for the metric can be integrated readily as $g=r^2$ and $\chi=\chi_0$.
Then, numeric solutions for $A_t$ and $\tac$, that depend on the value of the charge of the scalar, 
correspond to a probe spacetime filling brane in an $AdS_4$ spacetime, with action given by just the square root 
term in \eqref{eq.DBImodel}.
For finite values of the tension we integrate numerically the whole set of equations \eqref{eqs.fullequations} using
the IR asymptotic solution \eqref{eq.zeroTexpansion} (expanded up to a higher order in $r^{2\alpha}$) as a seed.

\begin{figure}[tb]
\begin{center}
\begin{subfigure}[b]{0.45\textwidth}
\includegraphics[width=\textwidth]{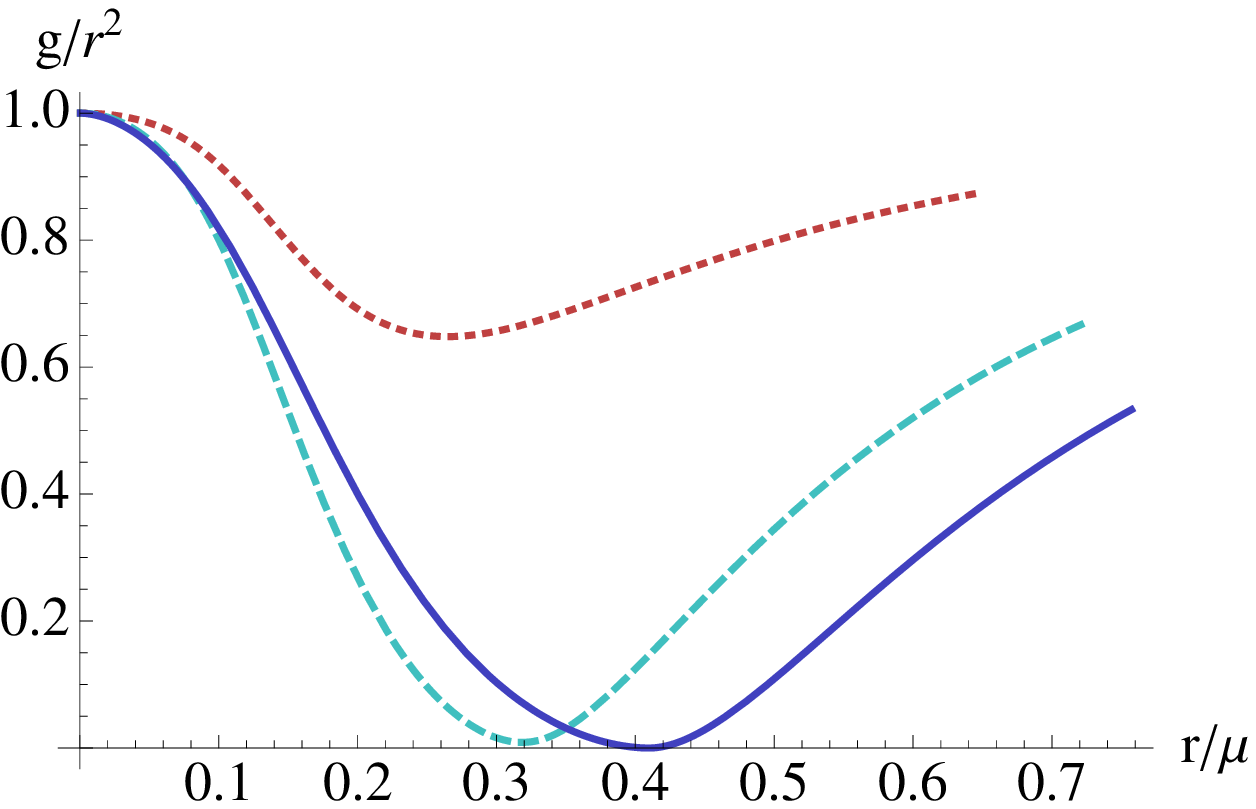}
\end{subfigure}
~
\begin{subfigure}[b]{0.45\textwidth}
\includegraphics[width=\textwidth]{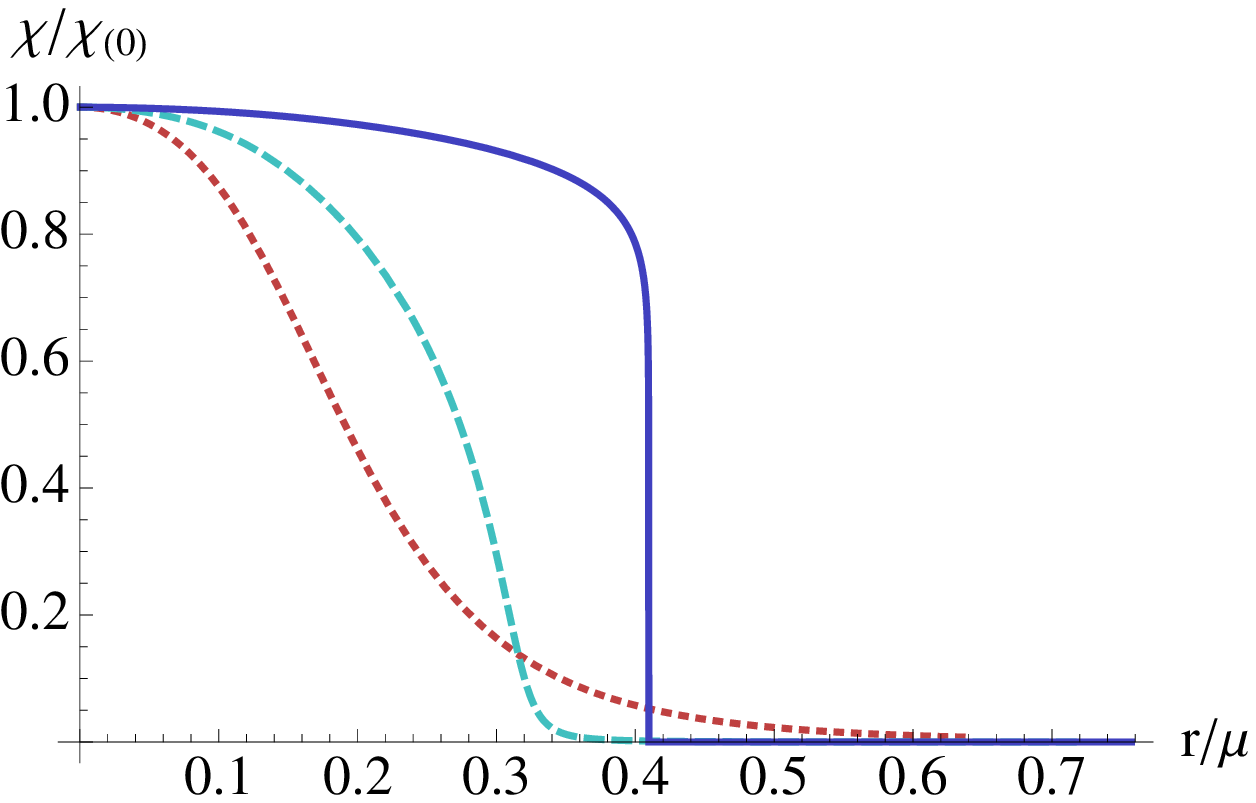}
\end{subfigure}\\[4mm]

\begin{subfigure}[b]{0.45\textwidth}
\includegraphics[width=\textwidth]{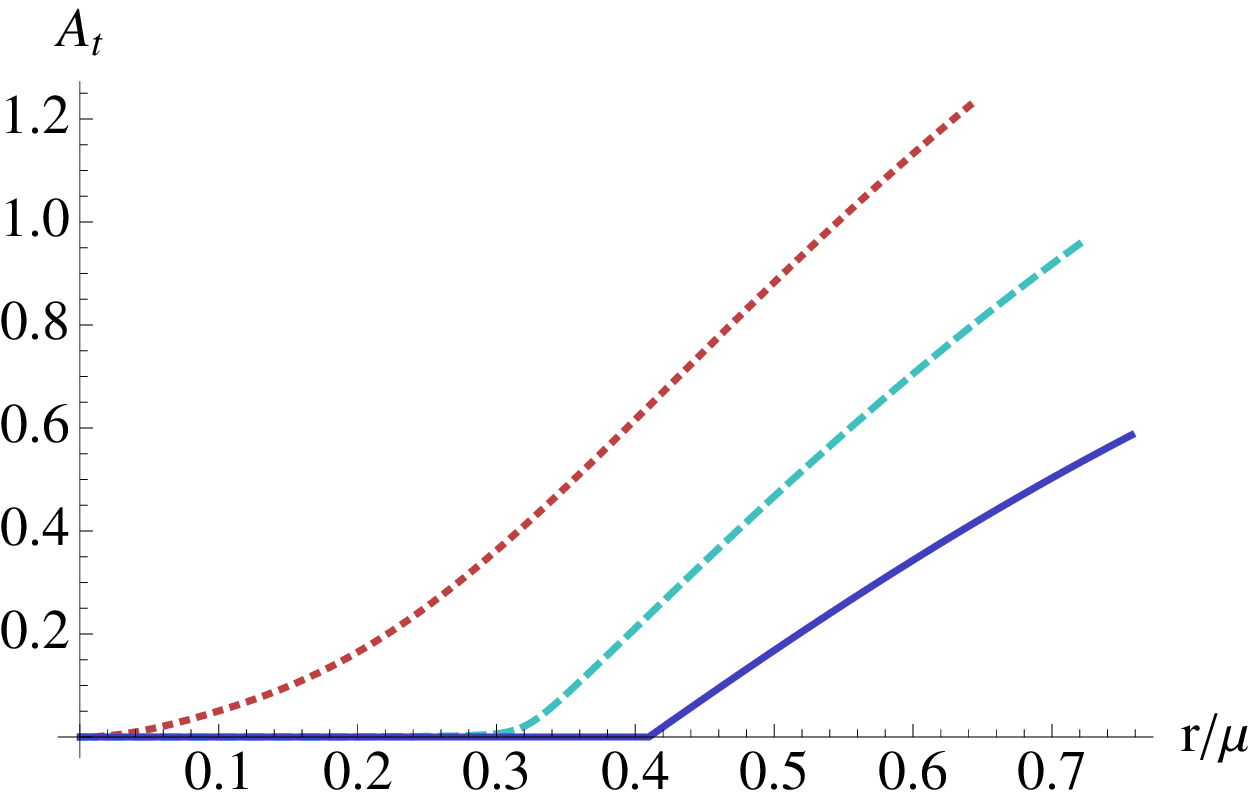}
\end{subfigure}
~
\begin{subfigure}[b]{0.45\textwidth}
\includegraphics[width=\textwidth]{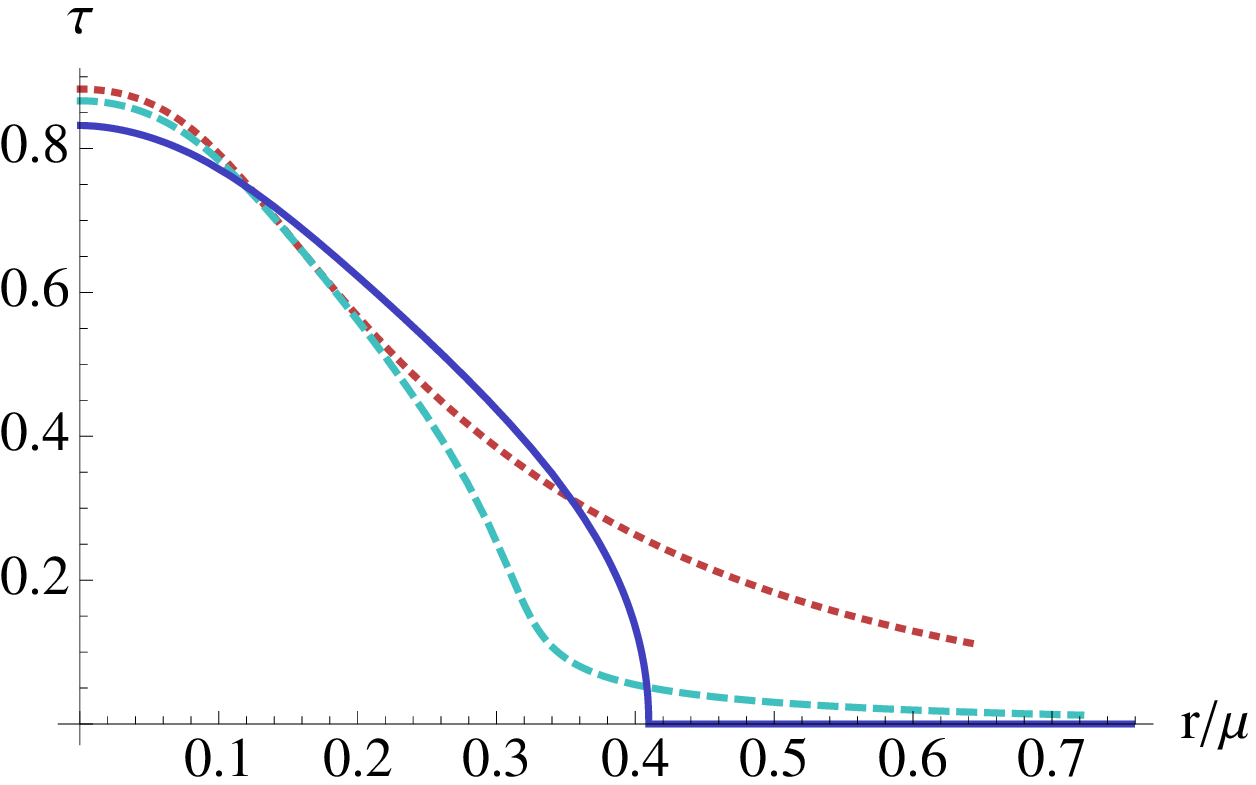}
\end{subfigure}
\caption{Plots of the functions in our ansatz as a function of $r/\mu$ at zero temperature, $m^2\,L^2=0$, $q\,L=2$. 
The different curves correspond to tensions $\tension\,L^2=1/2$ (red, dotted), $\tension\,L^2=3/2$ (cyan, dashed) 
and $\tension\,L^2=\tension^c\,L^2=7/3$ (blue, solid).} \label{fig.zeroTexamples}
\end{center}
\end{figure}
In figure \ref{fig.zeroTexamples} three examples of the numerical integration are shown for a charge $q\,L=2$
(in each case $\alpha$ has been tuned to a value such that the source term of the scalar is 
$|\tac_0(r=10^7)|\lesssim 10^{-12}$).
In view of these results we shall next study how the system behaves as one approaches the critical values of the parameters
for which the instabilities studied in section \ref{sec.instabilities} occur.

First, remember that for vanishing tension the system becomes a probe in $AdS_4$, and indeed at low values of $\tension\,L^2$  
the value of $g/r^2$ is close to the $AdS_4$ value $g/r^2\sim1$ everywhere.
As the tension is increased a dip appears in $g(r)$, with the minimum of the dip having a positive value. 
We know from the analysis in section \ref{sec.instabilities} that there exists a critical value of the tension $\tension^c$
beyond which the condensed phase does not exist anymore.
When this critical tension is reached, $\tension\,L^2=\tension^c\,L^2$, the minimum of $g$ occurs at $(g/r^2)|_{r=r_0}=0$, 
signaling the appearance of an extremal black hole with horizon radius $r_0$. 
Moreover, this numerical solution coincides accurately with the extremal black hole normal phase solution, given 
in eqs.~\eqref{eq.normalphase}, for $r>r_0$.

Regarding the scalar $\tac$, for the critical case $\tension\,L^2=\tension^c\,L^2$ the scalar  is different from zero
only inside the horizon, $r<r_0$, whereas for lower values of the tension its profile is non trivial everywhere, 
indicating (a) the existence of a condensate for $\tension\,L^2<\tension^c\,L^2$ related to the normalizable mode 
of the scalar at the boundary, and (b) that in the condensed phase the charge at zero temperature is extended
in the bulk of the spacetime, whereas for the normal phase it is contained in the extremal black hole
\cite{Gubser:2009cg,Horowitz:2009ij}.

The metric function $\chi$ behaves similarly to the scalar field $\tac$ (notice that in figure \ref{fig.zeroTexamples} 
we have scaled its value at the origin to one for presentational purposes), being zero outside the horizon for the critical
case, and extending into the bulk for lower values of the tension.
Generically, the value $\chi(0)$ at the origin grows with the tension up to a maximum at $\tension\,L^2=\tension^c\,L^2$ 
(and it is zero for $\tension\,L^2=0$).

Finally, the gauge field profile vanishes for $r<r_0$ when $\tension\,L^2=\tension^c\,L^2$, and extends down to  
$r=0$, vanishing there, for lower values of the tension.

As a further check of our numerics, we have verified that the values  for the condensate obtained at zero temperature
coincide with the  values one gets by extrapolating the low temperature behavior observed, for example, in figure 
\ref{fig.changeinordercondensate}. 
Actually, the whole numerical solution at finite temperature asymptotes to the $T=0$ one as one reduces the temperature.

\begin{figure}[t]
\begin{center}

\begin{subfigure}[b]{0.475\textwidth}
\includegraphics[width=\textwidth]{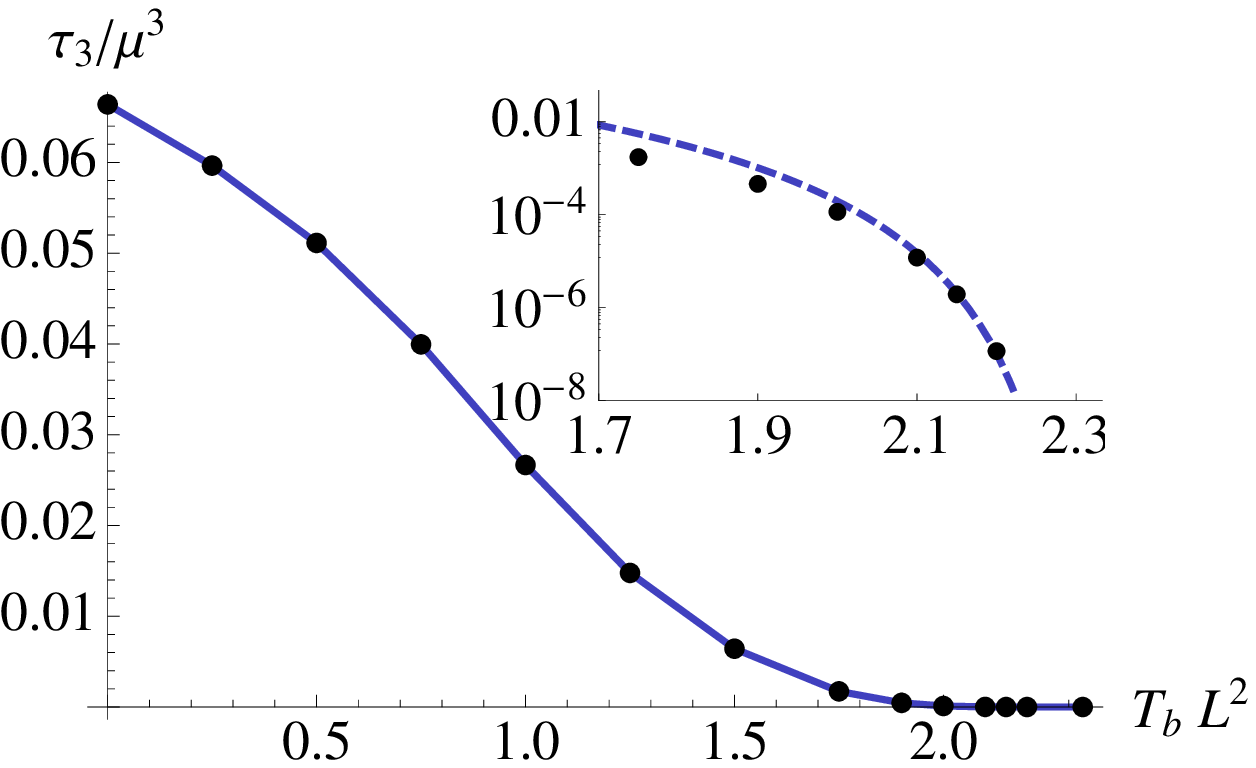}
\caption{$q\,L=2$\label{fig.zeroTcondensateqis2}}
\end{subfigure}
~
\begin{subfigure}[b]{0.475\textwidth}
\includegraphics[width=\textwidth]{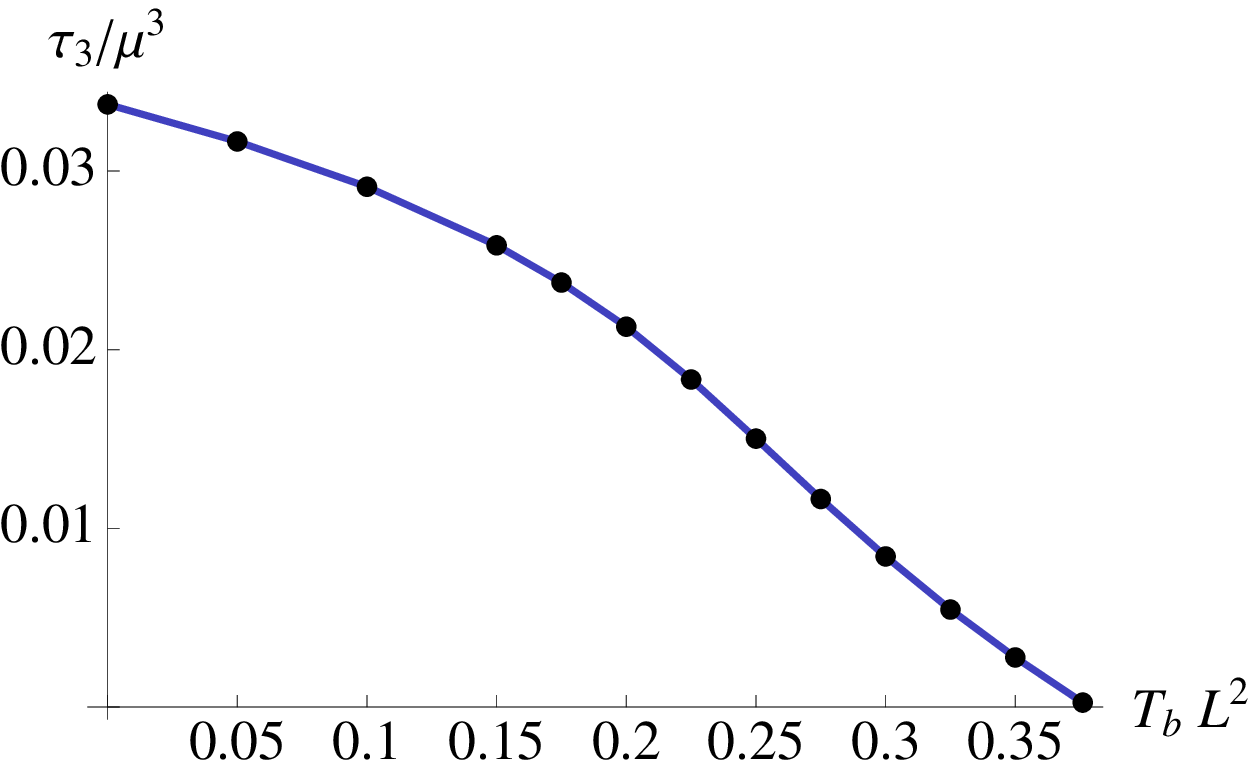}
\caption{$q\,L=3/2$\label{fig.zeroTcondensateqis1p5}}
\end{subfigure}
\caption{Condensate at zero temperature for $m^2\,L^2=0$. The inset in the left-hand side plot is a zoom on the 
exponential tale. The dashed line corresponds to expression \eqref{eq.BKT} for the holographic BKT phase transition 
with a fitted value of $\Lambda_{\text IR}$. The solid line  connects linearly the numerical data.} \label{fig.ZeroTemp}
\end{center}
\end{figure}

To conclude this section let us corroborate that, as stated in section \ref{sec.instabilities}, when the quantum
phase transition occurs due to the violation of the BF bound of the near horizon $AdS_2$ spacetime, this phase
transition is a holographic BKT (continuous) transition. While this is not the case for
the second kind of instability discussed in section \ref{sec.instabilities}

In a BKT phase transition, the behavior of the condensate at zero temperature as the critical point 
$\tension\, L^2=\tension^c\,L^2$ is approached must be given by the expression \eqref{eq.BKT}.
We shall then compute the value of the condensate at $T=0$ as a function of the tension, in the vicinity of the critical
point  \eqref{eq.maxtension}.
We present our results in figure \ref{fig.zeroTcondensateqis2}, where on the inset we provide a zoomed-in version of 
the exponential tale. The dashed line there corresponds to the  prediction of eq. \eqref{eq.BKT} after a one parameter fit
using the position of the rightmost point shown in the plot.
Instead, in figure \ref{fig.zeroTcondensateqis1p5} we provide the same plot for a value of the charge for which
the instability at zero temperature occurs away from the horizon (\emph{i.e.}, it is the second kind of instability 
discussed in section \ref{sec.instabilities}). 
It is clear from this figure that there is no exponential holographic BKT behavior. This was expected since the instability
is not caused by the violation of the BF bound in $AdS_2$, and the quantum phase transition is actually second order.

\section{Phase diagram for the model with a $\Delta=2$ operator}\label{sec.relevant}

In this section in order to study the theory where the condensing operator has dimension $\Delta = 2$ we 
take the mass of the scalar field to be $m^2\,L^2=-2$.
With this choice the scalar field behaves asymptotically as
\begin{equation}
\tac  \simeq  \frac{\tau_1}{r} +  \frac{\tac_2}{r^2}  + \cdots \ , \qquad (r\to\infty)
\end{equation}
Both $\tau_1$ and $\tau_2$ are normalizable modes, and hence two different quantizations are possible \cite{Klebanov:1999tb}.
The two theories correspond to the cases where $\tac_1$ ($\tac_2$) is dual to the source of a dimension
$\Delta=2$ ($\Delta=1$) operator and $\tac_2$ ($\tac_1$) is dual to the VEV.

For the remaining functions in our ansatz, the equations of motion imply that in the UV ($r\to\infty$)
\begin{subequations}
\begin{align}
g & \simeq r^2 \left[ 1 +  \frac{\tension}{2}  \frac{\tac_1^2}{r^2} + \frac{{\cal E}}{r^3} + \cdots \right] \ , 
\label{eq.gUVd2}\\
\chi & \simeq\chi_0 +   \frac{\tension}{2}  \frac{\tac_1^2}{r^2} +  \frac{4\,\tension}{3}  \frac{\tac_1\tac_2}{r^3}+ \cdots \ , \\
A_t & \simeq \bar \mu + \frac{\bar Q}{r} + \cdots \ .
\end{align}
\end{subequations}
From now on we pick the quantization in which the operator has dimension $\Delta=2$, and correspondingly
$\tac_1$ is interpreted as the source.
For the scalar to condense spontaneously we impose $\tac_1=0$.
In this case the expressions for the physical quantities given in eqs.~\eqref{eq.freeenergy} and
\eqref{eq.numericexpressions} hold, with the value of $\cal E$ to be read from the UV asymptotics of $g$ given
in eq.~\eqref{eq.gUVd2}.
The condensate density of the dual operator obtained from \eqref{eq.condensate} reads 
\begin{equation}
2\kappa^2 \frac{\langle {\cal O}_\tac \rangle}{V_2} = \tension\, \tac_2 \ .
\end{equation}

\begin{figure}[tb]
\begin{center}
\includegraphics[width=0.8\textwidth]{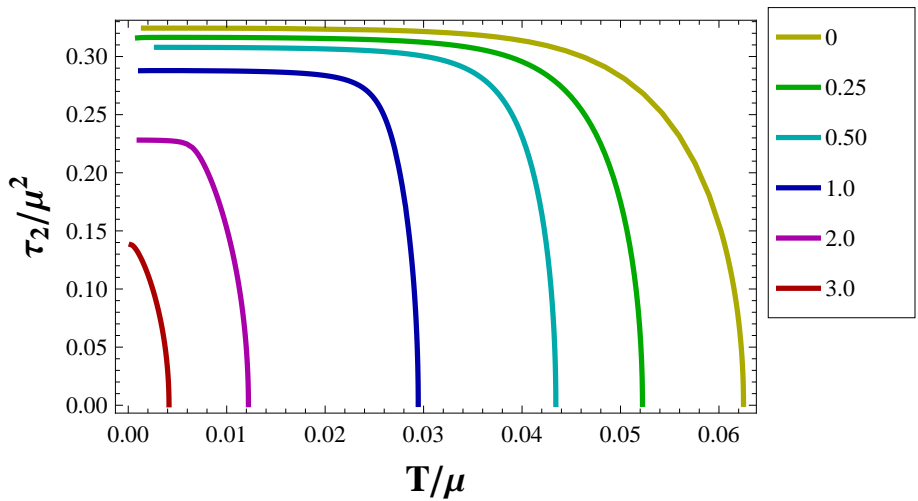}
\caption{(Color online) Values of the condensate  as a function of the temperature for ${m^2\,L^2=-2}$ and $q\,L=2$.
Each curve corresponds to a different value of the tension (see inset). We have checked that all the phase transitions
are of second order.
\label{fig.masminus2condensate}}
\end{center}
\end{figure}
The same procedure as described in the previous section is employed to construct the numerical solutions corresponding to 
the condensed phase of the system at finite temperature.
In order to illustrate these solutions, a plot of the condensate versus the temperature is shown in figure 
\ref{fig.masminus2condensate}, while for the phase diagram in the $q\,L$ -- $\tension\,L^2$ parameter space we refer the 
reader to figure \ref{fig.BFboundmmin2}.
Recall that in that figure the solid line corresponds to holographic BKT transitions at zero temperature, whereas the dashed 
one corresponds to $T=0$ second order phase transitions.
Moreover, we have found that, as opposed to the case in the previous section, for a relevant operator of dimension 
$\Delta=2$ the phase transition to the condensed phase is always of second order at finite temperature.

\begin{figure}[tb]
\begin{center}
\begin{subfigure}[b]{0.45\textwidth}
\includegraphics[width=\textwidth]{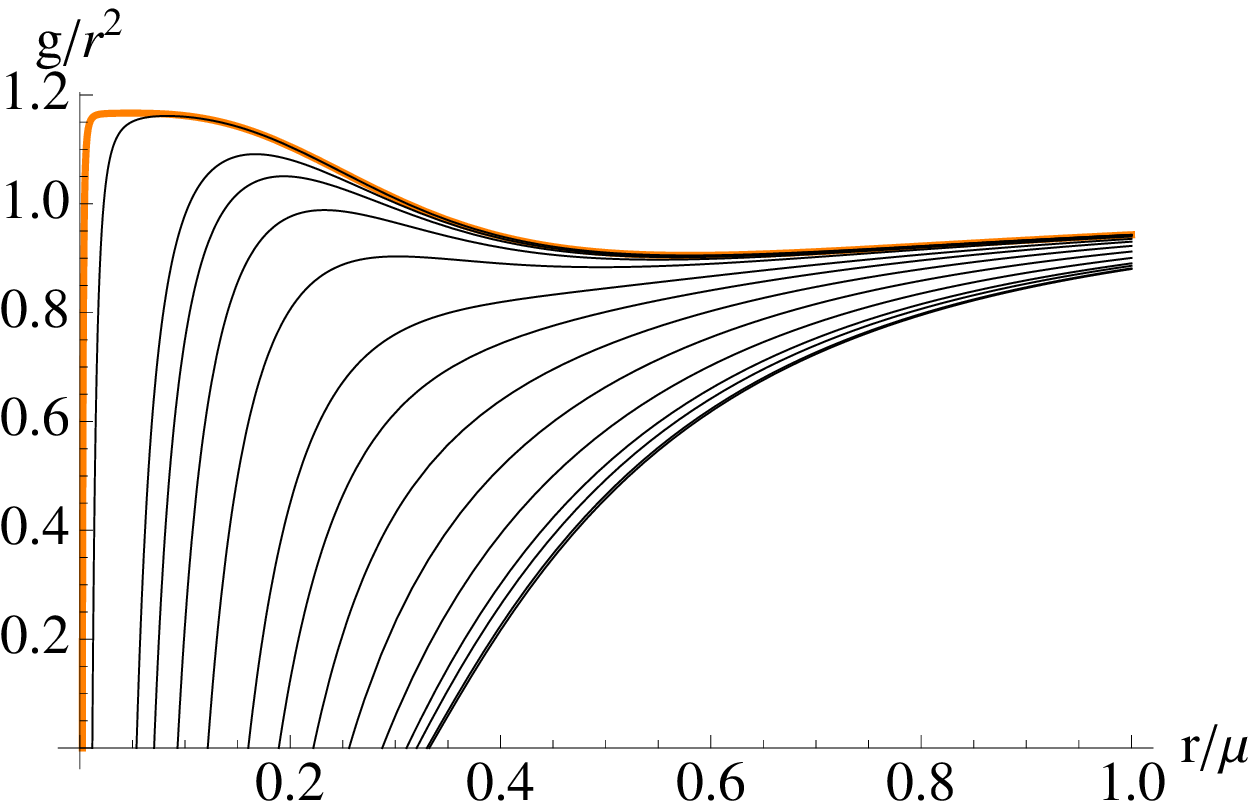}
\end{subfigure}
~
\begin{subfigure}[b]{0.45\textwidth}
\includegraphics[width=\textwidth]{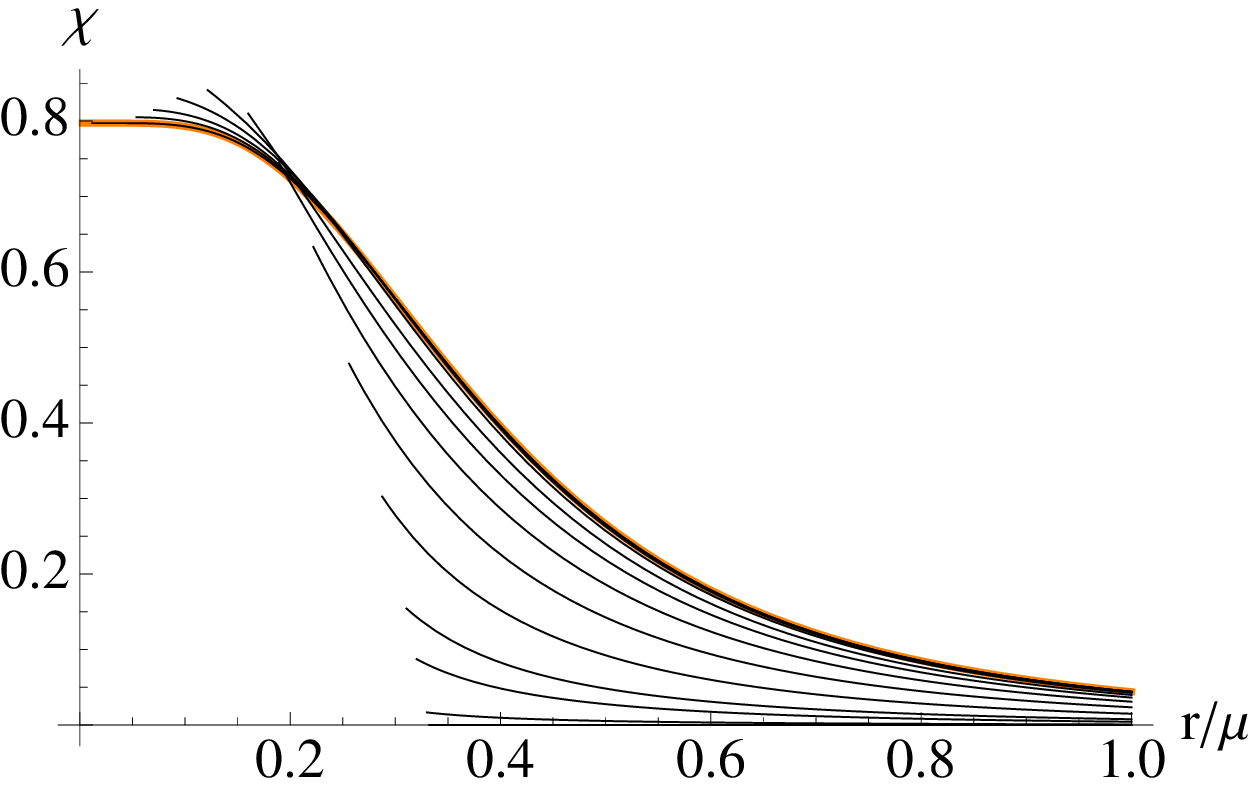}
\end{subfigure}\\[4mm]

\begin{subfigure}[b]{0.45\textwidth}
\includegraphics[width=\textwidth]{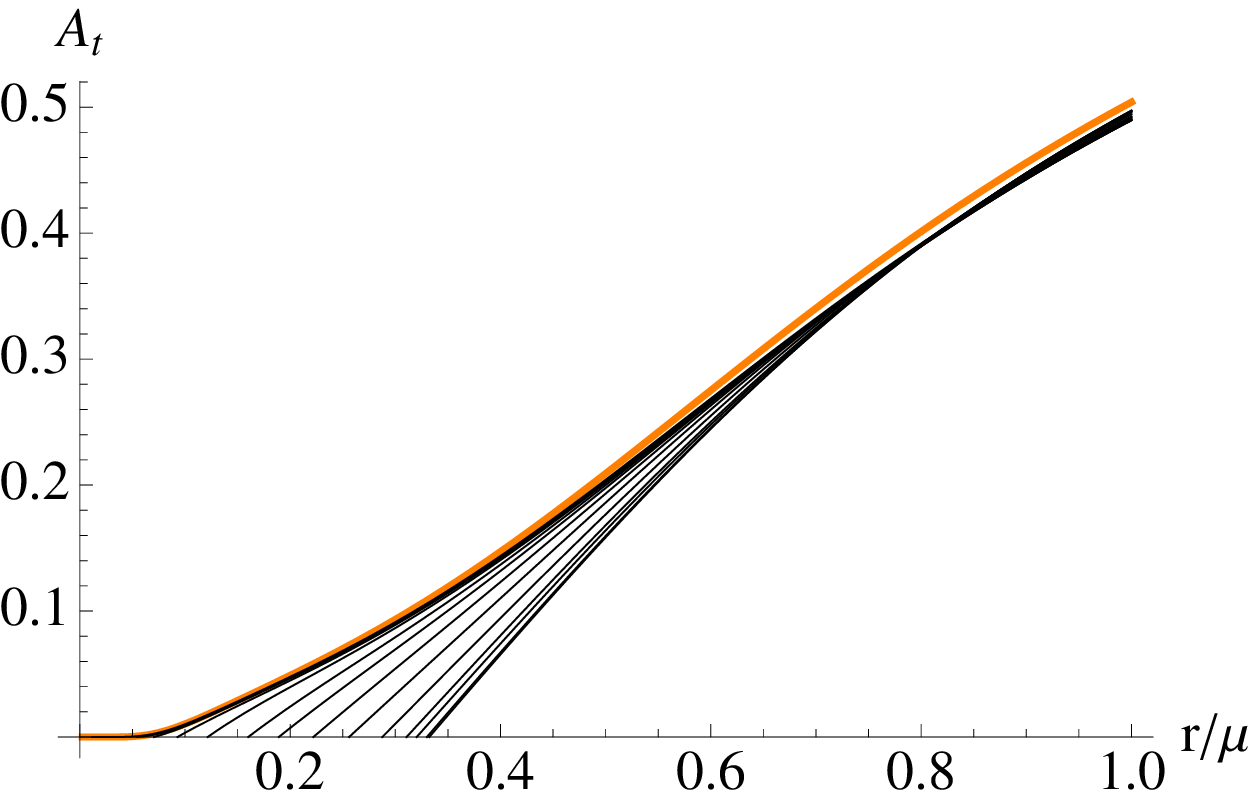}
\end{subfigure}
~
\begin{subfigure}[b]{0.45\textwidth}
\includegraphics[width=\textwidth]{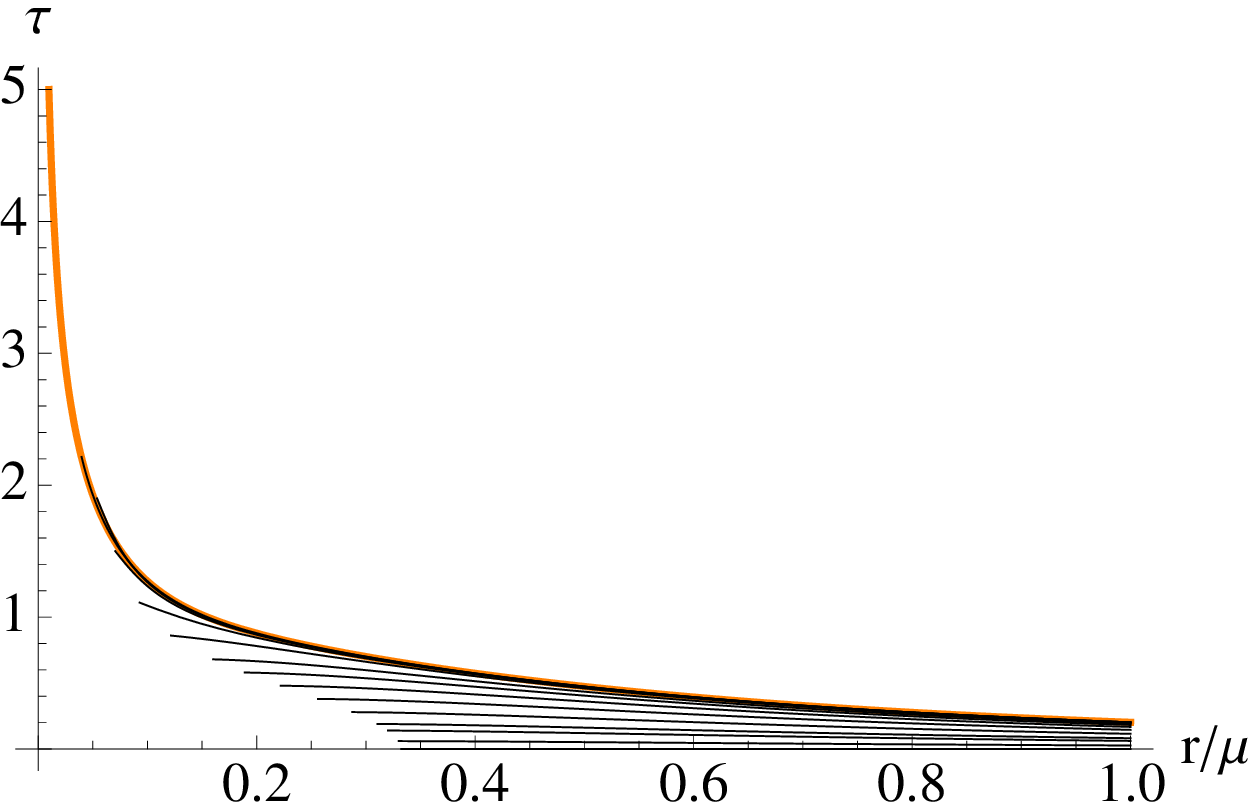}
\end{subfigure}
\caption{Plots of the functions in our ansatz as a function of $r/\mu$ at $m^2\,L^2=-2$, $q\,L=2$, and $\tension \, L^2=1$. The different curves correspond to decreasing temperatures from $T=T_c$ to $T=0.015\,T_c$  (thicker, orange line).} \label{fig.lowTexamples}
\end{center}
\end{figure}

For the case at hand, { \emph i.e.} $m^2\,L^2=-2$, we have not constructed the solutions corresponding
to the ground state (zero temperature) of the condensed phase. However, by studying the behavior of our
solutions at very low temperature one can gain intuition on the nature of the ground state.
%Let us first describe 
The result is presented in figure \ref{fig.lowTexamples}.
First, notice from the plot of the function $g/r^2$ that the radius of the horizon (where $g(r_h)=0$) decreases, in units of the chemical 
potential, as the temperature is decreased. 
Precisely at the horizon $A_t(r_h)=0$, while $\chi(r_h)$ and $\tac(r_h)$ go to constants; as required by the boundary 
conditions. 
%As for the metric functions $g/r^2$ vanishes at the horizon while $\chi$ goes to a constant there.
However, the most interesting feature of these plots is  the region that appears at small radius  in the $T\to0$ limit. In this region both
$g/r^2$ and $\chi$ stabilize to a constant, hinting to an $AdS_4$ geometry for the ground state.
%which point towards  This region is 
%Moreover, the metric function $\chi$ seems to stabilize to a constant near the horizon as the temperature 
%is decreased further. 
The scalar field $\tac$  diverges towards the horizon, while the gauge field 
$A_t$ is suppressed exponentially as $r\to0$.
This picture is confirmed by the equations of motion \eqref{eqs.fullequationsans} in the small radius limit. 
Indeed, at leading order in the IR,
one can find a solution where the matter fields $\tac$ and $A_t$ give exponentially suppressed corrections to the geometry. The metric becomes 
that of $AdS_4$ with radius $(1+\tension\,L^2/6)L$, while the scalar diverges as $r^{-4/(6+\tension\,L^2)}$ when $r\to0$, whereas the
gauge field vanishes exponentially $A_t\sim \exp\left(-r^{-8/(6+\tension\,L^2)}\right)$.
We leave for the future the construction of the  zero temperature solution for the whole range of the radial coordinate, which is expected to be a domain wall interpolating between two $AdS_4$ geometries \cite{future}.
Finally, notice that when $r/\mu\to\infty$, the gradient of $\tac$ approaches a constant that remains basically unchanged 
for many (low) values of the temperature, corresponding to the flat parts of the curves in figure 
\ref{fig.masminus2condensate}.
% For the case at hand, { \emph i.e.} $m^2\,L^2=-2$, we have been unable to find the solutions corresponding
% to the ground state (zero temperature) of the condensed phase.
% However, in order to gain some intuition about the nature of the ground state we studied the behavior of our
% solutions at very low temperature. The result is presented in figure \ref{fig.lowTexamples}.
% First, notice from the plot of the function $g/r^2$ that the radius of the horizon decreases, in units of the chemical 
% potential, as the temperature is decreased. 
% At precisely the horizon $A_t(r_h)=0$, while $\chi(r_h)$ and $\tac(r_h)$ go to constants; as required by the boundary 
% conditions. Moreover, the metric function $\chi$ seems to stabilize to a constant near the horizon as the temperature 
% is decreased further. Instead, the scalar field $\tac$ seems to diverge at the horizon in the $T\to0$ limit. 
% On the other hand, when $r/\mu\to\infty$, the gradient of $\tac$ approaches a constant that remains basically unchanged 
% for many (low) values of the temperature, corresponding to the flat parts of the curves in figure 
% \ref{fig.masminus2condensate}.

Regarding the entropy density, at low temperatures this quantity behaves as dictated by the $AdS_4$ geometry, namely it vanishes quadratically with the temperature.

To finish this section let us comment on the region for which the condensed phase does not exist
for intermediate values of the tension $\tension\,L^2$
% only for large or small values of the tension $\tension$, with the normal phase being stable down to zero temperature at intermediate values of $\tension$ 
(as it can be seen in figure \ref{fig.BFboundmmin2} this happens for values of the charge above 
$q\,L=1/2$ and below $q\,L\sim1.936$).
From the discussion above, the zero temperature transition that occurs at small values of the tension 
(along the dashed line of Fig.~\ref{fig.BFboundmmin2}) is of second order,
whereas the $T=0$ transition at large values of the tension (continuous line in Fig.~\ref{fig.BFboundmmin2}) is of the 
holographic BKT type. 
Moreover, the typical critical temperature for large values of the tension is several orders of magnitude smaller 
than the chemical potential, and it grows slowly as the tension is increased (this last fact is true also for values of 
the charge such that the condensed phase exists for any value of the tension, namely those above $q\,L\sim1.936$).
%\jt{was this true? then what happens for a charge like q=2 when you go to larger tnesions?}

\section{DC and optical conductivities}\label{sec.conducs}

From the phenomenological point of view it is of great interest to calculate the retarded current-current correlators,
and in particular the conductivities associated to the different currents.
Holographically we achieve this by studying the set of fluctuation modes that couple to the spatial perturbation
of  $U(1)$ gauge fields. 
In the UV we require our configuration to realize an electric field of constant modulus and frequency $\omega$
along the $x$ direction, while in the IR we impose ingoing boundary conditions at the horizon \cite{Son:2002sd}.
The conductivity is then computed as the quotient between the normalizable and non-normalizable modes of the spatial, 
zero momentum perturbation of the $U(1)$ field, divided by $i\, \omega$.

Recall that, as discussed in section \ref{sec.model}, in our setup there are two gauge fields corresponding
to the axial and vectorial $U(1)$s. In the absence of chemical potential along the $U(1)_V$ the fluctuations corresponding
to switching on an electric field in either of the $U(1)$s do not mix with each other, and hence we can study them separately.
We start the discussion with the vectorial conductivity. Therefore we consider the following perturbation of the $U(1)_V$
gauge field:
\begin{equation}
\label{eq.vecfluc}
\delta V = e^{-i \omega t}v_x(r) dx\ .
\end{equation}
To write the  Lagrangian quadratic in the fluctuations we must consider the most general version of the model,
for which the DBI term is given by eq. \eqref{eq.Sen}.
The fluctuation \eqref{eq.vecfluc} decouples from any other mode, and the relevant piece of the action is
then given by
\begin{equation}\label{eq.vectorialfluc}
{\cal L}_V^{(2)} \sim \frac{\tension}{2\kappa^2} \, \frac{1}{g_{\rm eff}^2(r)} \, \sqrt{ -\det \, \cM} \, 
F^V_{\mu\nu}\cW^{(\nu\alpha)}F^V_{\alpha\beta}\cW^{(\beta\mu)} \ ,
\end{equation}
with $\cM$ and $\cW$ defined in eqs.~(\ref{eq.curlw} -\ref{eq.wcomp}), and $F^V$ the field strength associated to
the $U(1)_V$ field. The effective, radial-dependent coupling reads
\begin{equation}\label{eq.effectivemaxwellcoupling}
\frac{1}{g_{\rm eff}^2(r)} = V(\tac)\, \frac{\sqrt{-g}}{\sqrt{-\cM}} \ ,
\end{equation}
and is proportional to the potential and the quotient of the metric and open string metric determinants.
The resulting equation of motion reads
\begin{equation}
\label{eq.vx}
v_x'' + \partial_r \log \left[\frac{1}{g_{\rm eff}^2}\,\sqrt{-\det \cM} \,\cW^{xx}\, \cW^{rr}\right] v_x' - 
\omega^2 \frac{\cW^{tt}}{\cW^{rr}} v_x = 0 \ .
\end{equation}
Since translation invariance is unbroken in our model we expect that charge will not dissipate in time, 
leading to a delta peak at zero frequency in the conductivity (equivalently, a $1/\omega$ pole in the imaginary part).
However, in the Lagrangian \eqref{eq.vectorialfluc} there is no mass term, which implies the absence of such a delta peak.
The reason for this is that the charge density associated to the $U(1)_V$ field is exactly zero in our 
setup,\footnote{Notice that in all our results in this section the chemical potential will be always that
along the $U(1)_A$.} and 
therefore the electric current, which is then due to pair creation only, carries no net momentum.

We will as usual read the AC conductivity from the asymptotic behavior of $v_x(r)$. For large $r$ the solution of
\eqref{eq.vx} behaves as
\begin{equation}
\label{eq.vxasympt}
 v_x = v^{(0)}+{v^{(1)}\over r} + o(r^{-2})\ .
\end{equation}
We will solve \eqref{eq.vx} numerically imposing ingoing boundary conditions
at the horizon, and then read the conductivity as
\begin{equation}
\label{eq.sigmacdef}
\hat \sigma(\omega) = {v^{(1)}\over i\omega\, v^{(0)}}\,.
\end{equation}
Notice that the numerically obtained quantity $\hat \sigma$ coincides with the conductivity up to a normalization
\begin{equation}
\sigma^V(\omega) = \frac{\tension}{2\kappa^2} \hat \sigma(\omega) \ .
\end{equation}
In particular $\lim_{\omega\to\infty} \hat \sigma=1$.

As for the DC conductivity in the $U(1)_V$ sector, isotropy and the effective Maxwell action \eqref{eq.vectorialfluc} 
allow us to follow the procedure of \cite{Iqbal:2008by} and express the DC conductivity  in terms of background functions 
evaluated at the horizon. Both for the normal and condensed phases the result can be written as
\begin{equation}
\sigma_{\rm DC}^V = \frac{\tension}{2\kappa^2} \frac{V(\tac_+)}{\sqrt{1-A_t'(r_+)^2}} \ ,
\label{eq.sigdcv}
\end{equation}
where $\tac_+$ is the value of the scalar at the horizon $r=r_+$. Notice that this expression becomes even simpler 
in the normal phase where the scalar vanishes and therefore $V(\tac_+)=1$

Let us see what we can learn about $\sigma_{\rm DC}^V$ in the condensed phase for the two cases analyzed in this paper.
In the model with $m^2\, L^2=0$, \emph{i.e.}, that with a marginal operator, the potential is just $V=1$, and
thus $\sigma_{\rm DC}^V$ is determined solely by the value of $A_t'$ at the horizon. Notice that $A_t'(r_+)$ vanishes
in the zero temperature limit (see the IR expansion \eqref{eqatzeroTIR}), and hence at $T=0$ the DC conductivity is a 
non zero constant. 

\begin{figure}[tb]
\begin{center}
\includegraphics[width=0.6\textwidth]{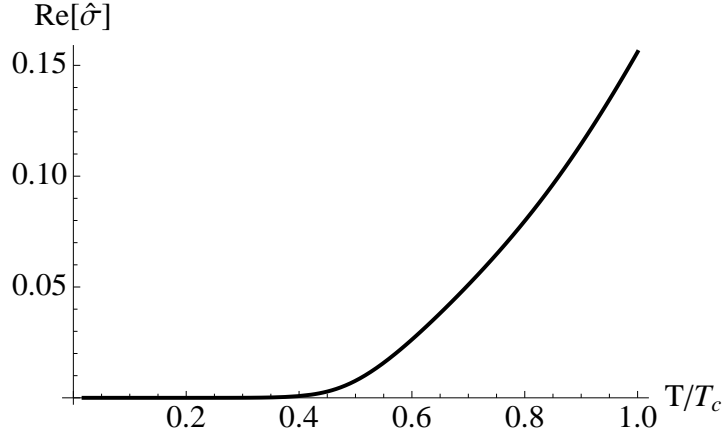}
\caption{DC conductivity of the vectorial current for  $m^2\,L^2=-2$, $q\,L=2$, and $\tension \, L^2=0.25$.} 
\label{fig.DCsigmavectorial}
\end{center}
\end{figure}

For the model with $m^2\, L^2=-2$, the DC conductivity of the $U(1)_V$ current takes the form
\begin{equation}
\label{eq.sigdcvm2}
\sigma_{DC}^V = \frac{\tension}{2\kappa^2} \frac{e^{-2\tac_+^2}}{\sqrt{1-A_t'(r_+)^2}} \ .
\end{equation}
Interestingly, as shown in figure \ref{fig.lowTexamples}, as $T\to 0$, $\tac_+$ diverges while $A_t'(r_+)\to0$,
 hence $\sigma_{DC}^V$ is suppressed as $e^{-2\tac_+^2}$.
Therefore, in this sector (for which the charge density is zero), the system behaves as an 
unconventional insulator.\footnote{Note that a similar feature was displayed in \cite{Mefford:2014gia} where in the absence 
of charge density the model exhibited an exactly vanishing DC conductivity (see also \cite{Baggioli:2014roa}).}
We provide an example of this feature in figure \ref{fig.DCsigmavectorial}, where 
we observe that below a certain temperature the DC conductivity is suppressed.
This suppression is an effect of the effective metric that governs the fluctuations of the vectorial field
(see eq. \eqref{eq.vectorialfluc}). This effective metric is sensitive to the value of the condensed scalar, 
and even when we have not broken the $U(1)_V$ symmetry we observe a pseudogap 
(the conductivity is small but not exactly zero) in the corresponding conductivity.
This is a direct effect of our DBI model.

We shall now focus on the 
conductivity associated to the axial $U(1)_A$ field. We must then consider the perturbation
\begin{equation}
\delta A = e^{-i \omega t}a_x(r) dx\ ,
\end{equation}
which is, as usual, coupled to the fluctuation $\delta g_{tx}$ of the metric due to the finite chemical potential 
associated to the $U(1)_A$ symmetry.
However, a constraint relating $\partial_r \delta g_{tx}$ to $a_x$, arising from the Einstein equations, allows us to 
decouple the $a_x$ mode completely and solve a simple linear differential equation of second order.\footnote{This is 
not unfamiliar in the contest of holographic superfluids, 
see for example \cite{Hartnoll:2008kx} for explicit expressions in a simpler setup.}
The effective Lagrangian for the axial perturbation is then given by \eqref{eq.vectorialfluc} with the substitution
$F_V\to F_A$ and the addition of a mass term. The resulting equation of motion for $a_x(r)$ takes the form
\begin{equation}\label{eq.axialfluc}
a_x'' + \partial_r \log \left[ \frac{1}{g_{\rm eff}^2}\,\sqrt{-\det \cM} \,\cW^{xx}\, \cW^{rr}\right] a_x' 
- \omega^2 \frac{\cW^{tt}}{\cW^{rr}} a_x + \frac{Y(r)}{{4 \left(e^{-\chi} \,g - 2 q^2\, A_t^2\, \tac^2\right)}} \, 
a_x = 0 \ ,
\end{equation}
with the mass term given by the following  expression
\begin{align}
Y  & =  2\, q^2\, \tac^2  \left[r\,e^{-\chi} \left(1+2\, g\, \tac'^2 - e^\chi \left(  A_t'^2+2\,q^2\, A_t^2\, \tac^2 \, g^{-1} \left(1+2\, g\, \tac'^2 \right) \right) \right) +2\,\tac\,  A_t' \,A_t^{3/2} \partial_r \left( \frac{r}{\tac\, \sqrt{A_t}} \right) \right]  \nonumber \\
& \qquad + \tension \, V(\tac) \, \frac{r\, {A_t'}^2}{ \sqrt{ 1+2\, g\, \tac'^2 - e^\chi \left(  A_t'^2+2\,q^2\, A_t^2\, \tac^2 \, g^{-1} \left(1+2\, g\, \tac'^2 \right) \right)  } } \ .
\end{align}
The UV expansion of $a_x$ is of the same form as \eqref{eq.vxasympt} and thus to compute the conductivity we solve
the equation \eqref{eq.axialfluc} numerically with infalling boundary conditions at the horizon, 
and read the conductivity using eq. \eqref{eq.sigmacdef}.
Notice that $Y(r)$ now implies the presence of a delta peak at zero frequency in the $U(1)_A$ conductivity.
Whenever we talk about the DC conductivity in this case we refer to the finite part that can be read from the zero 
frequency limit of the optical conductivity.

\begin{figure}[tb]
\begin{center}
\begin{subfigure}[b]{0.45\textwidth}
\includegraphics[width=\textwidth]{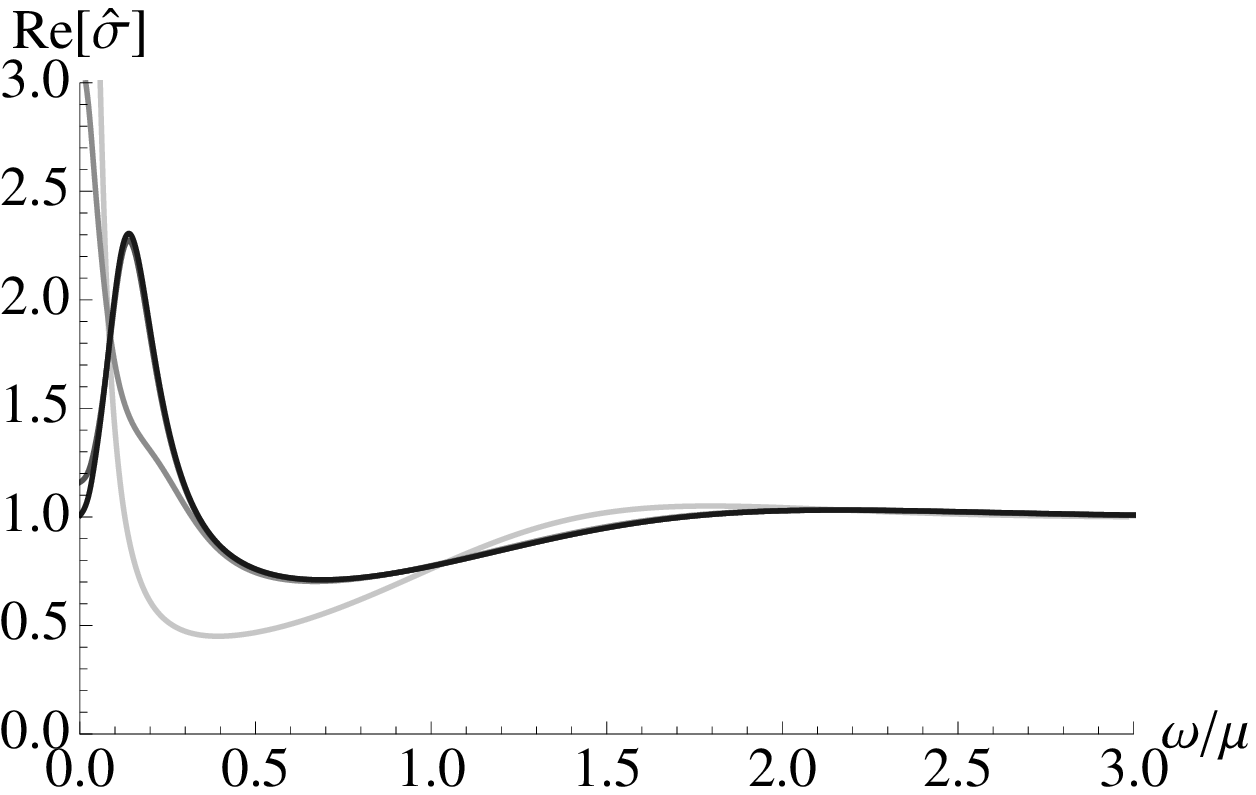}
\caption{Vectorial field.} \label{fig.vecsigmam0}
\end{subfigure}
~
\begin{subfigure}[b]{0.45\textwidth}
\includegraphics[width=\textwidth]{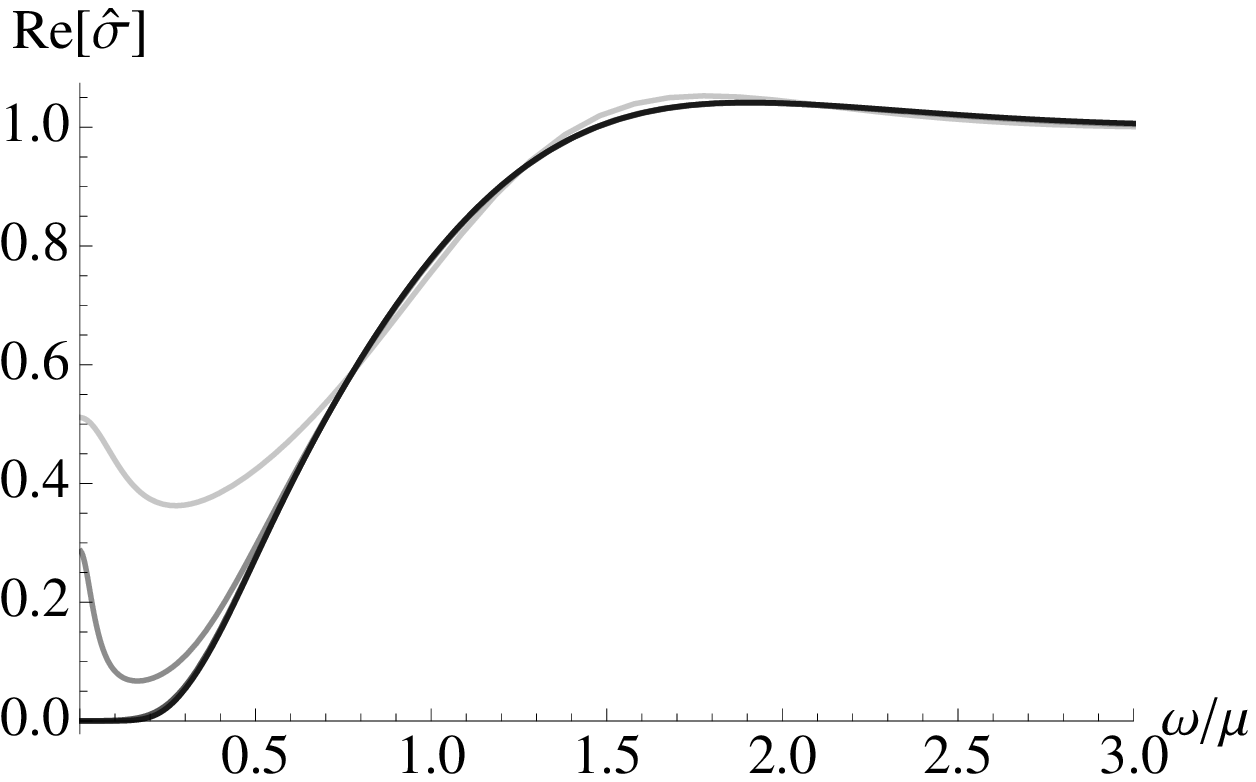}
\caption{Axial field} \label{fig.axialsigmam0}
\end{subfigure}
\caption{Real part of the (a) vectorial conductivity and (b) axial conductivity in the condensed phase as a 
function of the frequency for $m^2\,L^2=0$, $q\,L=2.5$ and $\tension \, L^2=0.15$. The line in the plot is darker the 
lower the temperature is, taking values from $T/T_c=\{1,\, 0.81,\, 0.41,\, 0.12\}$ (the two last values are almost
indistinguishable in the plot).} 
\end{center}
\end{figure}

Let us now present and discuss the plots with our numerical results for the optical conductivities.
First, in figure \ref{fig.axialsigmam0} we plot the real part of the axial conductivity for the condensed phase 
of the model with $m=0$.
These results are very similar to what was obtained for the conductivity of the minimal holographic superconductor in
\cite{Hartnoll:2008kx}. Namely, the AC conductivity approaches a constant at large frequencies (as determined by 
the $AdS_4$ asymptotics of
the solution), while at low frequency a pseudogap appears when the temperature is low enough.
In this figure we also observe that the AC axial conductivities for temperatures smaller than 
$T\sim 0.4 T_c$ remain basically invariant. The reason is that for low values of the tension, and
temperatures below $\sim 0.4 T_c$, the dynamics is well approximated by that of the zero temperature solution
discussed in section \ref{subsec.qpt} (whose geometry corresponds to an $AdS$ domain wall)
This can be argued by looking at figure \ref{fig.changeinordercondensate} where one can see that at 
$T\sim 0.4 T_c$ the condensate has already stabilized at the corresponding zero temperature value.
In figure \ref{fig.vecsigmam0} we plot the real part of the vectorial conductivity also for the condensed phase 
of the model with $m=0$. As discussed above there is no delta peak in this sector, and moreover no pseudogap appears either.
Again the conductivity practically reaches the zero temperature curve for temperatures below $\sim 0.4 T_c$.
Moreover as $T\to0$, the DC conductivity tends to the $AdS_4$ conformal value $T_b/(2\kappa^2)$,  as predicted 
by eq.~\eqref{eq.sigdcv}. Notice that in figure  \ref{fig.vecsigmam0} the two lines with larger temperature go to a 
constant at zero frequency, although this is not visible since we have truncated the axis for presentational purposes.

\begin{figure}[tb]
\begin{center}
\begin{subfigure}[b]{0.45\textwidth}
\includegraphics[width=\textwidth]{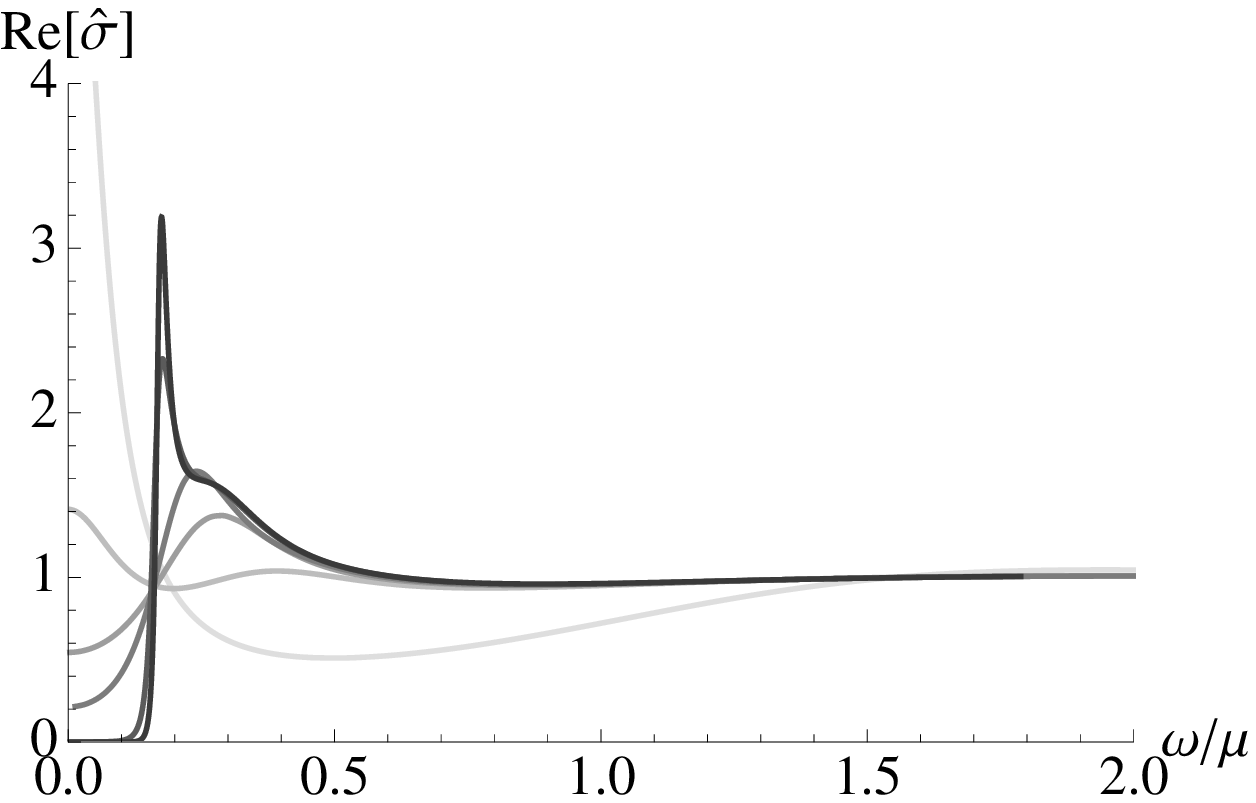}
\caption{Vectorial field} \label{fig.vectorsigmamm2}
\end{subfigure}
~
\begin{subfigure}[b]{0.45\textwidth}
\includegraphics[width=\textwidth]{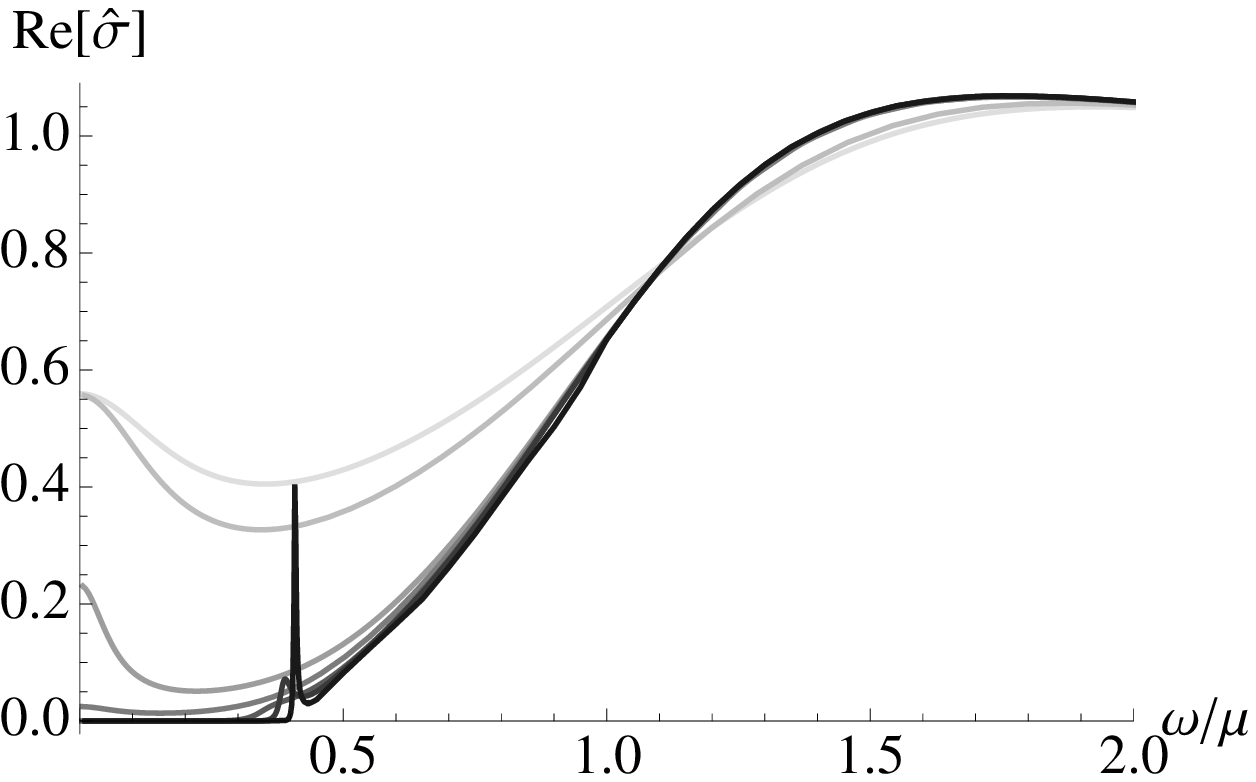}
\caption{Axial field.} \label{fig.axialsigmamm2}
\end{subfigure}
\caption{Real part of the (a) vectorial conductivity and (b) axial conductivity in the condensed phase as a function 
of the frequency for  $m^2\,L^2=-2$, $q\,L=2$ and ${\tension \, L^2=0.25}$. The line in the plot is darker the lower the 
temperature is, taking values from $T/T_c=1$ (lighter) to $T/T_c=0.09$ (darker).\label{fig.sigmamm2}} 
\end{center}
\end{figure}

In figure \ref{fig.vectorsigmamm2} we provide the  $U(1)_V$ optical conductivity in the condensed phase for
the case with $m^2\,L^2=-2$. Let us first emphasize again that no delta peak appears in the zero frequency limit. Indeed,
as we discussed below eq. \eqref{eq.sigdcvm2} the DC conductivity is finite, and moreover is highly suppressed as the 
temperature approaches zero.
Next, also for low temperature,  a pseudogap is clearly visible at low frequencies. The existence of this pseudogap is a 
consequence of the behavior of the tachyon in the IR. As we have seen in figure \ref{fig.lowTexamples}, the tachyon 
diverges towards the horizon, and hence the factor $V(\tau)$ appearing in the effective coupling 
\eqref{eq.effectivemaxwellcoupling} has the effect of a soft wall suppressing the spectral density at low frequencies 
as in \cite{Atmaja:2008mt}.
Since the AC conductivity satisfies a sum rule, 
the suppression of the $\omega=0$ conductivity has to be compensated by an increase of the spectral density at finite 
values of the  frequency. This shift of spectral weight is confirmed by the numerical solution; 
at low temperatures a peak forms at the end of the pseudogap.
%a frequency set by the value of the chemical potential. 
 After that peak there is a region with approximately 
constant conductivity, followed by a decrease towards the value determined by the asymptotic $AdS_4$ geometry at
large frequencies.

\begin{figure}[tb]
\begin{center}
\includegraphics[width=0.6\textwidth]{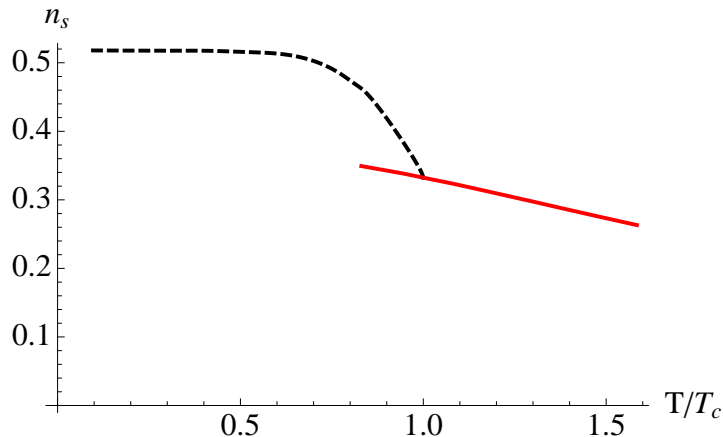}
\caption{Pole of the imaginary part of the axial conductivity $n_s$ for the $m^2\,L^2=-2$ case. $n_s$
results from  a fit of Im$[\sigma]$ to the function $n_s/(\omega/\mu)$, for values of $\omega/\mu\lesssim0.1$.
The solid red line
corresponds to the normal phase, while the dashed black line to the broken phase.} 
\label{fig.ns}
\end{center}
\end{figure}

In figure \ref{fig.axialsigmamm2} we present the real part of the AC conductivity for the axial current.
As discussed above, a delta peak at zero frequency is present both for the condensed and normal phases.  This delta peak 
is visible via the dispersion relations through the associated $1/\omega$ pole in the imaginary part of the conductivity.
One can  confirm that the weight of the delta function receives a new contribution (proportional
to the superfluid density) once the system is in the condensed phase. Indeed, as shown in figure \ref{fig.ns}
(where we study the evolution of the residue of the pole of the imaginary part of the conductivity as a function of
temperature) one observes a continuous change in the behavior of the zero-frequency pole as the system enters the 
superfluid phase.
We also observe that at low frequencies and low enough temperature the AC conductivity presents a pseudogap. 
Notice that the width of this region is largely independent of the temperature, $\omega_g \sim 0.42\mu$.

To end this section, let us insist on  a feature of figure \ref{fig.sigmamm2}. 
At very low temperatures (about 10$\%$ the critical temperature) a sharp peak becomes visible at the end of the 
pseudogap region. This peak becomes higher and narrower as the temperature is decreased, and we have checked that it 
appears quite generically when varying parameters of the theory,  being easier to observe for low values of the tension
$\tension\,L^2$. We comment on its possible significance in the next section.

\section{Discussion}\label{sec.conclus}

In this work we have constructed a holographic superfluid with matter transforming in the fundamental representation of 
the gauge group. 
The main novelty of our model 
comes from the use of
the tachyonic action describing a spacetime filling
$D3$--$\overline{D3}$ in an asymptotically $AdS_4$ BH geometry.
%The main difference between our model and others considered before is the tachyonic action describing a spacetime filling $D3$--$\overline{D3}$ in an asymptotically $AdS_4$ BH geometry. 
The setup contains a scalar field, the tachyon, charged under the $U(1)_A$ included in the global 
$U(1)\times U(1)$ symmetry supported by the $D3$--$\overline{D3}$,  but neutral under the diagonal $U(1)_V$. 
We have turned on a finite chemical potential corresponding to the $U(1)_A$, and found that below some critical
temperature the tachyon condenses breaking $U(1)\times U(1)\rightarrow U(1)_V$, and hence realizing a 
superfluid phase transition. We have focused on two classes of theories, one where the operator dual to the tachyon 
is marginal ($\Delta=0$), and the other one corresponding to a relevant operator with $\Delta=2$.

We have shown that there are two gravitational types of instability that drive the transition to a condensed phase, 
one of them giving rise to a holographic BKT phase transition at zero temperature, whereas the other leads to a second 
order one. 
For the case with a marginal operator we have constructed the zero temperature solution and shown that  the low temperature 
dynamics is governed in the IR by a conformal fixed point with the same central charge as the fixed point governing the UV.

Once the mass of the tachyon is fixed, there are two further  parameters in the model; these are the tension of the branes 
(which controls the amount of backreaction), and the charge of the tachyon. We have studied the phase
diagram of the system as a function of these two parameters plus the temperature.
When the operator that condenses is marginal there are two different types of finite temperature phase transitions. 
Depending on the explicit values of the charge and tension, the finite temperature transition can be first or second order. 
In the former case there exist metastable  phases close to the critical point, whereas in the latter the value of the 
condensate provides an order parameter with mean field exponent.
When the operator that condenses has dimension $\Delta=2$ the phase transition at finite temperature is always second order. 

In the last part of this work we have studied the  AC and DC conductivities for our setup, both for the current associated 
to the $U(1)_V$ and the $U(1)_A$  symmetries. In the $U(1)_A$, the conductivity shows features very similar to those of 
the minimal holographic superconductor of ref.~\cite{Hartnoll:2008kx}. It displays a delta peak at zero frequency both in 
the normal and condensed phases, with the weight of the delta function receiving a new contribution when the system enters 
the broken phase. As expected, the axial optical conductivity presents a pseudogap at low temperatures. 
The $U(1)_V$ DC conductivity is instead finite both in the normal and broken phases. This is due to the absence of charge
in that sector, for which only the pair produced charged carriers contribute to the conductivity. We have
produced a neat expression for this DC conductivity in terms of the horizon data, showing that, for the case where
the operator dual to the tachyon has $\Delta=2$, as the temperature is lowered the DC conductivity goes to zero faster 
than exponentially, and thus the system behaves similarly to an insulator in this sector.
As for the optical conductivity of the $U(1)_V$ sector, a pseudogap appears at low temperatures for the theory with 
a $\Delta=2$ operator
due to the scalar potential acting as a soft wall in the IR for the fluctuations. As expected, no pseudogap is
observed for the theory with a marginal operator
(where the tachyon is massless, and therefore the potential is trivial). It would be interesting to investigate what 
are the effects of a non trivial $U(1)_V$ chemical potential on the conductivities.

Finally, let us comment on an interesting feature of the optical conductivities for the model with a $\Delta=2$ operator. 
At low temperature, and
for low values of the tension, there appears a sharp peak at the end of the pseudogap region. This peak is the signature of 
a quasinormal mode (QNM, a normalizable solution of the fluctuation equation)
%a mode with the asymptotic behavior \eqref{eq.vxasympt} with $v_0=0$)
located at a specific position in the complex frequency plane, $\Omega$. For larger temperatures the position of this 
QNM has a negative imaginary part $\text{Im} \left[\Omega\right]\sim -T$. When the temperature is lowered this mode comes 
closer to the real frequency axis, and a peak with a certain width and height appears in the spectral function; a 
quasiparticle. If eventually the mode becomes real, $\text{Im}\left[\Omega\right]=0$, the peak becomes a delta function 
located at $\omega=\Omega$, and corresponds to a particle of the spectrum of the theory with mass $M=\Omega$. 
It is then natural to conjecture that the theory with a relevant operator of dimension $\Delta=2$ has, at zero temperature 
in the condensed phase, a particle  precisely at $M=\Omega$.
%This would be the first mode of the spectrum, which is therefore gapped.
In principle more massive modes would be present as well, and be observable in the spectral function at extremely low 
temperatures, but we were not able to obtain enough control on the numerics to observe them. 
We plan to come back to this point in future work \cite{future}.

A second line of future research would consist in the study of unbalanced superconductors. 
In order to do so one needs switch on a chemical potential along the diagonal $U(1)_V$. 
Since the tachyon is not charged under this $U(1)_V$ one can interpret this second chemical potential as
an imbalance of populations in a mixture of two species \cite{Erdmenger:2011hp}-\cite{Amado:2013lia}.
%The tachyon is not charged under this $U(1)_V$, and therefore as in \cite{Erdmenger:2011hp}-\cite{Amado:2013lia} one could interpret this second chemical potential as a measure of the imbalance of populations in a mixture of two species \cite{imbrvw}. 
This is an interesting line of study, since at weak coupling the so called LOFF inhomogeneous condensed phase \cite{loff} 
is expected to appear \cite{imbrvw}. In this context, the appearance of first and second order phase transitions in the 
phase diagram of the theory with a marginal operator is already hopeful. As reviewed in \cite{Bigazzi:2011ak}, 
in the phase diagram of unbalanced superconductors there is a region, at large values of the imbalance, where the phase 
transition becomes first order, and it is in this region where the LOFF phase is predicted to appear. 
It would then be interesting to consider the model with $\Delta=0$ and values of the parameters close to the region in 
the phase diagram where the phase transition becomes first order (see figure \ref{fig.limittensions}), and study the effect 
of a non zero chemical potential along the diagonal $U(1)_V$. Notice that while the two $U(1)$s  in  \cite{Bigazzi:2011ak} 
did only interact with each other through their backreaction on the geometry, thanks to the non-linearity of the DBI action, 
interacting terms will be present in our setup. 

Finally, a more ambitious continuation of this work would be to consider generalizations of this model along the lines 
of \cite{ihqcdtandmu} with a non trivial dilaton and consequently more general potentials depending both on the tachyon 
and the dilaton. This approach could lead to an interesting phenomenology as in \cite{Gouteraux:2012yr} for 
Einstein-Maxwell theories.

%
%As argued in the previous section, for $\tension\, L^2\to0$ and zero temperature the model is basically a probe brane moving in AdS$_4$, with the scalar field diverging at the horizon. \jt{Quero volver ler o paper de elias et al. onde calculan cousas na probe}

\section*{Acknowledgements}
We would like to thank Francesco Bigazzi, Aldo Cotrone, Roberto Emparan, and Ignacio `C\'ampora' Salazar, for useful discussions.
D.A. was partially supported by the COST program `String Theory Universe' while working on this project.
J.T. is supported by the grants 2014-SGR-1474, MEC FPA2010-20807-C02-01, MEC FPA2010-
20807-C02-02, CPAN CSD2007-00042 Consolider- Ingenio 2010,  ERC Starting Grant
HoloLHC-306605, FPA2013-46570-C2-2-P, and by the Juan de la Cierva program of the Spanish
Ministry of Economy.
We thank the Mainz Institute for Theoretical Physics (MITP) for hospitality and partial support during the
completion of this work. D.A. thanks the FRont Of pro-Galician Scientists for unconditional support.

\end{document}